\RequirePackage[l2tabu, orthodox]{nag}
\RequirePackage{snapshot}

\documentclass[9pt,onecolumn]{extarticle}

\makeatletter
\if@twocolumn
  \usepackage[dvips,letterpaper,top=0.5in, bottom=0.5in, left=0.75in, right=0.5in,includefoot,heightrounded]{geometry}
\else
  \usepackage[dvips,letterpaper,margin=1in,includefoot,heightrounded]{geometry}
\fi

\usepackage{srcltx}

\usepackage{natbib}
\usepackage[russian,portuges,english]{babel}

\iflanguage{portuges}
    {\newcommand{\keywordname}{Palavras-chaves}}
    {\newcommand{\keywordname}{Keywords}}

\usepackage{amsmath}
\usepackage{amssymb,amsfonts}

\usepackage{abstract}

\usepackage{graphicx}
\usepackage[usenames,dvipsnames,svgnames,x11names]{xcolor}
\usepackage{subfigure}
\usepackage{ctable} %

\usepackage{booktabs}

\usepackage{setspace}
\usepackage{flushend}

\usepackage{hyperref}
\usepackage[normalem]{ulem}

\usepackage{enumerate}

\usepackage[noend]{algpseudocode}

\usepackage{listings}

\usepackage[short,12hr]{datetime}

\usepackage{dsfont}

       \usepackage{fouriernc}

\newcommand{\barma}{${\beta}$ARMA}
\newcommand{\tablesize}{\fontsize{9}{11}\selectfont}

\newcommand{\Z}{\mathds{Z}}

\makeatletter

\newcommand{\printtitle}{%
\makeatletter
\if@twocolumn

\twocolumn[%
  \maketitle
  \begin{onecolabstract}
    \myabstract
  \end{onecolabstract}
  \begin{center}
    \small
    \textbf{\keywordname}
    \\\medskip
    \mykeywords
  \end{center}
  \bigskip
]
\saythanks
\else
  \maketitle
  \begin{onecolabstract}
    \myabstract
  \end{onecolabstract}
  \begin{center}
    \small
    \textbf{\keywordname}
    \\\medskip
    \mykeywords
  \end{center}
  \bigskip
  \onehalfspacing
\fi
\makeatother
}

\author{%
B. G. Palm%
\thanks{%
Programa de P\'os-gradua\c{c}\~ao em
Engenharia de Produ\c{c}\~ao,
Universidade Federal de Santa Maria, RS, Brazil,
E-mail: brunagpalm@gmail.com}
\and
F. M. Bayer%
\thanks{%
Departamento de Estat\'istica and LACESM,
Universidade Federal de Santa Maria, RS, Brazil,
E-mail: bayer@ufsm.br}
\and
R.~J.~Cintra%
\thanks{%
Signal Processing Group,
Departamento de Estat\'istica,
UFPE, Brazil,
E-mail: rjdsc@de.ufpe.br}
}

\title{%
Prediction Intervals
	in the Beta Autoregressive Moving Average Model}

\newcommand{\myabstract}{%
In this paper,
we propose five
prediction intervals
for
the beta autoregressive moving average model.
This model is suitable for modeling and forecasting variables
that assume values in the interval $(0,1)$.
Two of the proposed prediction intervals are based
on approximations considering
the normal distribution
and
the quantile function of the beta distribution.
We also consider bootstrap-based prediction intervals, namely:
(i)~bootstrap prediction errors (BPE) interval;
(ii)~bias-corrected and
acceleration
(BCa)
prediction interval;
and
(iii)~percentile prediction interval
based on the quantiles of the
bootstrap-predicted
values
for two different bootstrapping schemes.
The proposed prediction intervals were evaluated
according to
Monte Carlo simulations.
The BCa prediction interval
offered the best performance
among the evaluated intervals,
showing lower coverage rate distortion and small average
length.
We applied our methodology
for predicting the water level
of the Cantareira water supply system in S\~ao Paulo, Brazil.
}

\newcommand{\mykeywords}{%
\barma\ model, prediction intervals,
bootstrap,  time series, forecast
}

\date{}

\begin{document}

\printtitle

\section{Introduction}

Prediction of future values is a
relevant problem in the statistical analysis
of time series
and it is
a subject of interest in different areas of
research
\citep{homburg2020,
	Hotta2016, hyndman2018forecasting, Vidoni2009}.
Forecasting is often based on point
estimate
derived from past values of the time series
under consideration~\citep{Cheung1998}.
However,
predictions
can also be presented
in interval form~\citep{Pascual2005}
by means of prediction intervals.
Such intervals can be used as
a
reference for comparing
different forecasting methods,
facilitating
the choice of the most suitable
approach
and
the exploration of different forecasting
scenarios~\citep{Chatfield1993,abberger2006kernel},
and are widely explored in the literature.
For instance,
prediction intervals for count time series
are discussed in~\cite{homburg2020}.
Prediction intervals
based on
a deep
residual network
and
residual-based bootstrap methods
for hydrological data
are derived
in~\cite{yan2021flow}
and~\cite{beyaztas2018},
respectively.
In~\cite{Hotta2016},
prediction intervals
for
univariate volatility models
with leverage effect
are introduced.
Prediction intervals
for long memory data
are presented in~\cite{rupasinghe2014obtaining}
and~\cite{rupasinghe2012asymptotic}.

Usually,
the construction of the prediction intervals
is based
on:
(i)~the assumption
that the
errors
of the model
are normally
distributed~\citep{Thombs1990,pascual2004,Clements2007,Li2011,Hotta2016}
and
(ii)~the knowledge of the
model
parameters~\citep{pascual2004,Hotta2016}.
However,
if such
conditions
do not
hold,
then
the nominal coverage of the prediction intervals
can be
unsatisfactory~\citep{Thombs1990}
and
prediction intervals can be derived based on
the bootstrap method,
with unknown parameters
and
without assuming any specific
distribution~\citep{Efron1979, Masarotto1990, Davison1997}.
This method
has
been
widely-employed
in time series modeling
and
it is capable of
providing
accurate
prediction intervals
and
inferential improvements,
as exemplified
in
bias correction of estimators~\citep{Palm2017,Kilian1998},
construction of confidence intervals for
model parameters~\citep{Spierdijk2015},
calculation of
Fourier coefficients
for the autocovariance function~\citep{dehay2018},
model selection criteria~\citep{bayer2015a,cavanaugh1997},
prediction intervals~\citep{staszewska2016improved},
and
hypothesis testing~\citep{Morley2009}.

A well-known and popular
time series
framework
for modeling and forecasting
is the autoregressive moving average
(ARMA)
model~\citep{Box2008,Oppenheim2009}.
However,
if
the data under analysis
do
not satisfy normality,
the
ARMA model may not be suitable~\citep{Box2008}.
For instance,
normal-based
forecasting
of
data
that are
restricted to the interval~$(0,1)$,
such as rates and proportions,
can result in
values outside the
specific
limits~\citep{Cribari2010,Pinheiro2011}.
This is due to the fact
that the
normal
distribution support
is the whole real line and not just the standard unit interval $(0,1)$.
Another approach for
modeling of
double-bounded
data
is the transformation of the variable of interest,~$y$.
However,
the results
from such transformation
should be interpreted in terms
of the
mean of the transformed variable
and
not
of the
mean of $y$~\citep{Cribari2010}.
That could be in disagreement
with
Jensen inequality,
because
the conditional mean of
the
transformed
variable
may
differ
from the transformation of the conditional mean
of the variable~\citep{grillenzoni1998,white1984}.

In this context,
the beta autoregressive moving average model~(\barma)~\citep{Rocha2009}
offers
a suitable
way
to model continuous data
from
the interval $(0,1)$.
This particular model
assumes
that the variable of interest
is
beta-distributed.
Because
the support of the beta distribution is
$(0,1)$,
the \barma~model
forecast data
is naturally
suitable
for continuous data in the interval $(0,1)$,
providing forecasts that are more consistent with
this type of
actual data.
The \barma~model has been
extended and explored in the literature.
In~\cite{pumi2021},
a dynamic model for double-bounded time series
with chaotic-driven conditional averages
based on the beta distribution is proposed and discussed.
An study of forecasting Brazilian mortality rates
due to occupational accidents
using the \barma~model
is presented in~\cite{melchior2020}.
The \barma~models
for long-range dependence and sazonal series
are derived in~\cite{pumi2019}
and~\cite{bayer2018beta}, respectively.
Finally,
bootstrap-based inferential improvements
and goodness-of-fit test
for hydrological time series modeling
based on the \barma~model
are discussed in~\cite{Palm2017}
and~\cite{scher2020goodness}, respectively.

In~\cite{Rocha2009},
it is
derived
an approach
for
point estimation,
large data record results,
and point forecast
for the
\barma~model.
Nevertheless,
to the best of our knowledge,
the problem of deriving prediction intervals
for the \barma~model
has not been
addressed in literature.
Thus,
we aim at filling this gap
with
a
first treatment
on
the subject.
Our goal in this
paper is
to present
a methodology
for deriving
prediction intervals
for the \barma~model.

One of the
proposed
methods is based on the
beta-distribution
quantiles.
Another proposed interval is based on the
method used in the
ARMA class of models,
considering the
normal
distribution and the variance of the
prediction error~\citep{Box2008}.
Both intervals are considered
without bootstrapping.
The approach
proposed in~\cite{espinheira2014}
for the construction of the
bias-corrected and acceleration~(BCa)
prediction interval
for the beta regression model
is adapted to the \barma~model.
Another
employed method
is
a modification
of the method
proposed in~\cite{Masarotto1990}
for
standard
autoregressive models.
We also considered an interval based on
the quantiles of the predictions
yielded from the residual and block bootstrapping schemes.
The methods
proposed in~\cite{Box2008},~\cite{Masarotto1990},
and~\cite{espinheira2014},
may suffer drawbacks,
such as
unguaranteed
assumptions
of
(i)~knowledge of parameters~\citep{Box2008};
(ii)~normality
when data may not be normal
\citep{Masarotto1990};
and
(iii)~independence for time series data~\citep{espinheira2014}.

The paper is organized as follows.
Section~\ref{s:barma}
discusses
\barma~model,
presenting
point forecasting and residual analysis.
In Section~\ref{s:ip},
the proposed prediction intervals
and different types of bootstrap resampling schemes are
showed.
Section~\ref{s:num} describes
Monte Carlo simulation
for
the proposed methods
and
numerical results are
examined.
To further evaluate the proposed method,
an application
to real-world measured data
is
detailed and discussed
in Section~\ref{s:aplicacao}.
Finally,
Section~\ref{s:con}
concludes the work.

\section{The Beta Autoregressive Moving Average Model}
\label{s:barma}

The \barma~model was proposed
in~\cite{Rocha2009}
and can be described as follows.
Let
$\{y_t\}_{t\in \Z}$
be a stochastic process
for which
$y_t\in(0,1)$ with probability 1,
for all $t\in\Z$
and
let
$\mathcal{F} _{t}=\sigma\{y_t,y_{t-1},\dots\}$
denote
the sigma-field generated by
past observations
up to time~$t$.
Assume that,
conditionally
on
the
information
set $\mathcal{F} _{t-1}$,
each
$y_t$ is distributed according to
$\operatorname{Beta}(\mu_t,\phi)$ distribution,
where
$\mu_t \in (0,1)$
is the conditional mean
and
$\phi >0$
is the precision parameter.
This
parametrization
of the beta distribution
is
commonly employed
in the context
of
time series and regression
models,
where
the goal is to model
the
mean of the response variable~\citep{Ferrari2004,Rocha2009}.
The conditional density of $y_t$,
given $\mathcal{F}_{t-1}$,
is~\citep{Rocha2009}:
\begin{align}
\begin{split}
\label{E:den}
f(y_t\mid\mathcal{F}_{t-1})
=
\frac{\Gamma(\phi)}{\Gamma(\mu_t\phi)\Gamma((1-\mu_t)\phi)}
y_t^{\mu_t\phi-1}
(1-y_t)^{(1-\mu_t)\phi-1}
,
\qquad
0<y_t<1,
\quad
t>0
.
\end{split}
\end{align}
The conditional mean and
conditional variance of
$y_t$
are given, respectively, by:
\begin{align*}
\operatorname{E}(y_t\mid \mathcal{F}_{t-1})
&= \mu_t,
\\
\operatorname{Var}(y_t \mid \mathcal{F}_{t-1})
&= \frac{V(\mu_t)}{1+\phi}
,
\end{align*}
where $V(\mu_t)=\mu_t(1-\mu_t)$.
For a fixed value of $\mu_t$,
the variance of $y_t$
decreases as $\phi$ increases~\citep{Rocha2009}.

The \barma$(p,q)$ model is defined by the following structure:
\begin{equation}
\label{E:modelo}
g(\mu_t)=\eta_t
=
\beta+\sum_{i=1}^{p}
\varphi_i g(y_{t-i})
+
\sum_{j=1}^q
\theta_jr_{t-j}
,
\end{equation}
where
$g:(0,1)\rightarrow\mathbb{R}$
relates $\mu_t$ to the linear predictor $\eta_t$~\citep{Benjamin2003},
being
a strictly monotone,
twice
continuously-differentiable
link function
as in the beta regression model~\citep{Ferrari2004};
$\beta \in \mathbb{R} $ is a constant;
$\varphi=(\varphi_1,\varphi_2,\ldots,\varphi_p)^\top$
and
$\theta=(\theta_1,\theta_2,\ldots,\theta_q)^\top$
are the autoregressive and moving average parameters, respectively;
and
$p$ and $q$ are the orders of the model~\citep{Rocha2009}.
The \barma~model,
proposed in~\cite{Rocha2009},
also
includes
a term that accommodates
covariates,
as discussed
in~\cite{Benjamin2003},
for a class of generalized ARMA models.

As suggested
in~\cite{Rocha2009} and~\cite{Benjamin2003},
the
moving average term of a generalized ARMA~model
can be defined in several ways,
such
as
the error
on the predictor scale, $r_t=g(y_t)-g(\mu_t)$,
or on the original  scale, $r_t=y_t-\mu_t$.
In this paper,
we adopt the error on the
predictor scale;
thus,
the model input
$r_t$
and the model output
are
on
the scale of $g(y_t)$.
Therefore,
the transfer function of the system,
which transforms $r_t$ into $y_t$,
is given by a dynamic linear
relationship that
can lead to a controllable system~\citep{Box2008}.
The conditional variance of the error term
is furnished by~\citep{Rocha2009}:
\begin{align}
\label{e:sigma}
\operatorname{Var}
\Big(
g(y_t)-g(\mu_t) \mid \mathcal{F}_{t-1}
\Big)
=
\sigma_t^2
\approx
\left[
g^{\prime} (\mu_t)
\right]^2
\frac{V(\mu_t)}{1+\phi}
.
\end{align}

The usual choices for
the link function~$g(\cdot)$
include
the logit, probit, and complementary log-log
functions~\citep{Koenker2009}.
The parameters of the \barma~model
can be
estimated by maximizing the logarithm of the conditional likelihood function.
More details on point estimation and
large data record results
for the \barma~model, conditional score function,
and conditional Fisher information matrix
are
available in~\cite{Rocha2009}
and~\cite{rocha2017e}.

\subsection{Forecasting}
\label{s:prev}

Let $y_1,y_2,\ldots,y_n$ be a sample from the $\beta$ARMA
model,
where
the conditional
maximum likelihood estimation (CMLE)
can
be employed
to obtain
estimates~$\widehat{\mu}_t$
of
the conditional
mean~$\mu_t$
and~$n$ is the sample size.
The estimates
$\widehat{\gamma}=
(\widehat{\beta},\widehat{\varphi}^\top,
\widehat{\theta}^\top,\widehat{\phi})^\top$,
for
$\gamma=(\beta,\varphi^\top,\theta^\top,\phi)^\top$,
can be applied
in
Eq.~\eqref{E:modelo},
providing~$\widehat{\mu}_t$
and
the predicted values,
$\widehat{y}_{n}(h)$,
$h$ steps forward:
\begin{align*}
\widehat{y}_{n}(h)
=
g^{-1}
\left(
\widehat{\beta}
+
\sum\limits_{i=1}^{p}
\widehat{\varphi}_i
g^\star(y_{n+h-i})
+
\sum\limits_{j=1}^{q}
\widehat{\theta}_j
r^\star_{n+h-j}
\right)
,
\end{align*}
where
\begin{align*}
g^\star(y_{n+h-i})
& =
\begin{cases}
g(\widehat{y}_{n}(h-i)),&\text{if $i< h$},\\
g(y_{n+h-i}), & \text{if $i\geq h$},
\end{cases}
\end{align*}
and
\begin{align*}
r^\star_{n+h-j}
&=
\begin{cases}
0,&\text{if $j< h$},\\
g(y_{n+h-j}) - g(\widehat{\mu}_{n+h-j}), & \text{if $j\geq h$}.
\end{cases}
\end{align*}
The diagnostic analysis of the fitted model
is based on
the assessment of the behavior of residuals.
The residuals are functions of the observed and predicted values
one step ahead~\citep{kedem2005}.
In particular,
the ordinary residuals
are given
by~\citep{kedem2005}:
\begin{align*}
\mathcal{R}_{1}(y_t,\widehat{\mu}_t)
=
g(y_t) - g(\widehat{\mu}_t)
.
\end{align*}
As an alternative to ordinary residuals,
we consider
the standardized
residuals~\citep{kedem2005}.
For
\barma~models,
the standardized
residuals in the predictor scale can be defined by:
\begin{align*}
\mathcal{R}_{2}(y_t,\widehat{\mu}_t)
&=
\frac{g(y_t)-g(\widehat{\mu}_t)}
{\sqrt{
		[g^{\prime}(\widehat{\mu}_t)]^2 V(\widehat{\mu}_t)/(1+\widehat{\phi})
	}
}.
\end{align*}
The radical term is
an approximate estimate
of
the standard deviation
furnished by the truncated Taylor series expansion
of the variance of $r_t$ in
Eq.~\eqref{e:sigma}.
Another option for residual evaluation
is given by~\citep{Espinheira2008b}:
\begin{align*}
\mathcal{R}_{3}(y_t,\widehat{\mu}_t)
&=
\frac{y_t^\ast-\widehat{\mu}_t^\ast}
{\sqrt{ \widehat{\nu}_t}}
,
\end{align*}
where
$\widehat{\nu}_t =
\psi^{\prime}(\widehat{\mu}_t\widehat{\phi})
+
\psi^{\prime}(\left(1-\widehat{\mu}_t\right)\widehat{\phi})$,
$y_t^\ast = \log \lbrace y_t / \left( 1 - y_t \right) \rbrace$,
$\widehat{\mu}_t^\ast = \psi (\widehat{\mu}_t \widehat{\phi} )
- \psi ( ( 1 -\widehat{\mu}_t ) \widehat{\phi} )$,
$\psi(\cdot)$
and
$\psi^{\prime}(\cdot)$
are the digamma and trigamma functions,
respectively~\citep{abramowitz1964}.

Zero mean and constant variance
of
the standardized
residuals
indicate
a
good
model
fit~\citep{kedem2005}.
The absence of autocorrelation,
partial autocorrelation,
and
conditional heteroscedasticity in the series of residuals
are expected~\citep{Box2008}.
Such conditions can be
verified
by means of the residual correlogram
or
Box-Pierce~\citep{BoxandPierce1970},
Ljung-Box~\citep{LjungandBox1978},
and
Lagrange
multiplier
tests~\citep{Engle1982}.

\section{Prediction Intervals}
\label{s:ip}

Although,
the estimate
of the mean square error of
a predicted value serves as an indicator
of the
forecast error
performance,
prediction
intervals can
offer probabilistic
interpretations~\citep{Guttman1970,Stine1982}.
Indeed,
a prediction interval is defined
by
the
upper and lower limits
associated
with a
prescribed
probability~\citep{Chatfield1993}
for each
future value~$y_{n+h}$,
$h=1,2,\ldots,H$,
where $H$ is the
desirable
forecast horizon.

In general,
the intervals have the following format:
\[
[LL_h;UL_h]
,
\]
where $LL_h$ and $UL_h$ are the lower and
upper prediction limits, respectively, for $y_{n+h}$.
When
the errors
are not normally distributed
and
the parameters are unknown,
the coverage rate
assumes values
that are lower than
the nominal
coverage~\citep{Thombs1990,Masarotto1990}.
The coverage of the intervals
is
expected
to
be
close to the
nominal level,
$1-\alpha$,
where $\alpha$ is
the significance level.
To circumvent such restrictive assumptions,
bootstrap methods~\citep{Efron1994, Davison1997}
have been considered for the construction of prediction intervals,
furnishing
less
distorted results.

The prediction intervals
proposed
for
the \barma~model
are presented in the following sections.
First we
introduce two
prediction intervals
without bootstrapping
based
(i)~on the Box \& Jenkins approach
and
(ii)~on the quantile function of the beta distribution,
which are
described in
Sections~\ref{s:bj} and~\ref{s:qbeta},
respectively.
Second
we propose
three bootstrap-based
prediction intervals
in
Sections~\ref{s:bpe},~\ref{s:bca},
and~\ref{s:percentil}.
The methods are
respectively
based on:
(i)~the bootstrap prediction error;
(ii)~the BCa
prediction interval
of the
beta regression model;
and
(iii)~the
quantiles of the
bootstrap
predicted values
from two different resampling schemes.

\subsection{Approximate Intervals Based on
	Normal
	Distribution (BJ)}
\label{s:bj}

Considering known parameters and normal moving averages errors,
we introduce a variation of
the Box~\&~Jenkins (BJ)
prediction interval
for the ARMA model~\citep{Box2008}.
The BJ prediction
interval for the \barma~model
is proposed as follows:
\begin{align*}
LL_h &=
g^{-1}
\left\{
g(\widehat{y}_n(h)) + z_{\alpha/2} \sqrt{\widehat{V}_n(h)}
\right\}
,
\\
UL_h &=
g^{-1}
\left\{
g(\widehat{y}_n(h)) + z_{1-\alpha/2} \sqrt{\widehat{V}_n(h)}
\right\},
\end{align*}
where $z_{\alpha/2}$ and $z_{1-\alpha/2}$
are the quantiles $\alpha/2$ and
$1-\alpha/2$ of the
standard normal distribution,
respectively.
The variance of the prediction
error for $h$ steps forward
$V_n(h)$
and its
estimate
$\widehat{V}_n(h)$
is
defined
in~\cite{Box2008}
as
$
V_n(h) =
(
1 + \Psi_1^2
+ \Psi_2^2 + \cdots + \Psi_{H-1}^2
)
\sigma^2_{n+h}
,
$
where
$\Psi_j= \varphi_1 \Psi_{j-1} + \cdots + \varphi_{p+d}\Psi_{j-p-d}-\theta_j$
with
$\Psi_0=1$,
$\Psi_m=0$ for $m<0$,
and
$
\widehat{V_n(h)} = ( 1 + \widehat{\Psi}_1^2 +
\widehat{\Psi}_2^2 + \cdots + \widehat{\Psi}_{H-1}^2 )
\widehat{\sigma}^2_{n+h},
$
where
$\widehat{\Psi}_j= \widehat{\varphi}_1
\widehat{\Psi}_{j-1} +
\cdots + \widehat{\varphi}_{p+d}
\widehat{\Psi}_{j-p-d}-\widehat{\theta_j}$
and
$\widehat{\sigma}^2_{n+h} = [ g^{\prime} (\widehat{y}_n(h)) ]^2
\frac{V(\widehat{y}_n(h))}{1+\widehat{\phi}}$.

\subsection{Approximate Interval Based on the
	Beta Distribution Quantiles (Qbeta)}
\label{s:qbeta}

Another prediction interval
introduced
in this work is
based on the beta distribution quantiles (Qbeta).
We suppose that
the
conditional probability of the future process
value $y_{n+h}$, given the previous information
up to instant~$n$,
follows
the beta distribution
with
mean~$\widehat{y}_n(h)$ and precision~$\phi_h$.
As
the forecasting horizon increases,
the variability of the prediction error also increases.
To measure
the loss of precision of the
predicted values,
we consider
the following
estimator
of
parameter $\phi_h$,
using
Eq.~\eqref{e:sigma},
as:
\begin{align*}
{\phi}_h &\approx \frac{[g^{\prime}({\widehat{{y}}_n(h))}]^2
	[{\widehat{{y}}_n(h)}(1-\widehat{y}_n(h))]- {V_n(h)}}{{V_n(h)}}.
\end{align*}
However,
the value of the quantity $V_n(h)$ is unknown
and
it needs to be estimated,
as shown in~\cite{Box2008}.
An
estimate
for $\phi_h$ is directly given by:
\begin{align*}
\widehat{\phi}_h &= \frac{[g^{\prime}(\widehat{y}_n(h))]^2
	[\widehat{y}_n(h)(1-\widehat{y}_n(h))]- \widehat{V}_n(h)}{\widehat{V}_n(h)}.
\end{align*}
As a consequence,
the limits of the Qbeta prediction interval for each $h$
are furnished by:
\begin{align*}
LL_h &= \mathfrak{u}_{n}^{ (\alpha/2)}(h), \\
UL_h &= \mathfrak{u}_{n}^{ (1-\alpha/2)}(h),
\end{align*}
where
$\mathfrak{u}^{\alpha}_{n}(h)$ is
the $\alpha$ quantile function of
the $\operatorname{Beta}(\widehat{y}_{n}(h),\widehat{\phi}_h)$ distribution.

However,
the prediction intervals
presented above may
show lower
coverage rates
when compared with
nominal levels,
if the
assumption of normality and knowledge
of the parameters
are
not satisfied~\citep{Thombs1990,Hotta2016}.
This issue can be addressed by means of
the
bootstrapping method,
as
it is
based on
the
empirical distribution
of the prediction errors
or
of the predicted values~\citep{Thombs1990}.
In the
next
section,
bootstrap-based prediction intervals
are presented.

\subsection{Bootstrap Prediction Errors Interval (BPE)}
\label{s:bpe}

In~\cite{Masarotto1990},
a bootstrap interval
based on the bootstrap prediction errors~(BPE)
is
introduced.
Such method aims at building
the empirical distribution of prediction errors,
considering
the
estimated
parameters
by the model and the
sample size of the time series.
By
(i)~adapting
the BPE interval
for the \barma~model
and
(ii)~considering the predicted value $g(\widehat{y}_{n}(h))$
and the prediction
residual $\mathcal{R}_{2}(\cdot,\cdot)$,
we have
the following limits:
\begin{align*}
LL_h^b& = g^{-1}
\left\lbrace
g(\widehat{y}_n(h)) +
\sqrt{\left[ g^{\prime}(\widehat{y}_n(h))\right]^2
	V(\widehat{y}_n(h))
	/(1+\widehat{\phi}_h)}
\mathcal{R}_{2(\alpha /2)}(y_{n+h}^b,\widehat{y}^b_n(h))
\right\rbrace, \\
UL_h^b&= g^{-1}
\left\lbrace
g(\widehat{y}_n(h)) + \sqrt{
	\left[ g^{\prime}(\widehat{y}_n(h))\right]^2
	V(\widehat{y}_n(h))/(1+\widehat{\phi}_h)}
\mathcal{R}_{2(1-\alpha /2)}(y_{n+h}^b,
\widehat{y}^b_n(h))
\right\rbrace
,
\end{align*}
where
$\mathcal{R}_{2(\alpha /2)}(y_{n+h}^b,\widehat{y}^b_n(h))$
and
$\mathcal{R}_{2(1-\alpha /2)}(y_{n+h}^b,\widehat{y}^b_n(h))$
are
the
$\alpha / 2 $
and
$1 - \alpha / 2 $
quantiles
of
the standardized residuals
$\mathcal{R}_{2}(y_{n+h}^b,\widehat{y}^b_n(h))$,
respectively,
for
$b=1, 2, \ldots, B$,
and the quantity~$B$  is
the number of bootstrap
replications.
The bootstrap
sample
$y_t^b$
is an occurrence of the beta distribution
with mean
$\widehat{\mu}_t$
and precision
$\widehat{\phi}$.
Each instantiation $y_{n+h}^b$
stems from
the $\operatorname{Beta}(\widehat{y}_{n}^b(h),\widehat{\phi}_h)$
distribution,
where $\widehat{y}^b_n(h)$ is defined by
Eq.~\eqref{e:futurosboot}.

\subsection{
	Bias-corrected and Accelerated Bootstrap
	Prediction Interval
}
\label{s:bca}

The BCa interval~\citep{Efron1994}
is
\color{black}
based on the percentiles
of the bootstrap distribution of
$g(y_n(h))$.
The percentiles depend on
$a$ and $z_0$,
where $a$
is the skewness
correction (acceleration)
and $z_0$
is the
bias correction
of
$g(y_n(h))$~\citep{Davison1997}.
Based on the distribution of the residuals
of the bootstrap estimates
$\mathcal{R}_{3}(y_{n+h}^b,\widehat{y}^b_n(h))$,
the following
estimate for~$z_0$
was introduced in~\cite{espinheira2014}:
\begin{align*}
\widehat{z}_0
=
\Phi^{-1}
\left\{
\frac{
	\#\!\left[
	\mathcal{R}_{3}(y_{n+h}^b,\widehat{y}^b_n(h)) < \mathcal{R}_m)
	\right]
}
{B}
\right\}
,
\end{align*}
where
$\Phi ^{-1}(\cdot)$
\color{black}
is the inverse of the cumulative standard normal distribution function~\citep{Efron1994},
$\mathcal{R}_m$ is the median of the residuals
$\mathcal{R}_{3}(y[n],\widehat{\mu}[n])$,
and
$\#(\cdot)$
returns the number of times that
its argument
holds true.
In~\cite{espinheira2014},
the following estimate for $a$
was supplied:
\begin{align*}
\widehat{a}= \frac{1}{6} \frac{\widehat{\omega}_n(h)}{\widehat{\upsilon}^{3/2}_n(h)},
\end{align*}
where
$\widehat{\omega}_n(h)=\widehat{\phi}^3_h \lbrace \psi^{\prime\prime} (\widehat{y}_n(h)\widehat{\phi}_h) - \psi^{\prime\prime} ((1-\widehat{y}_n(h))\widehat{\phi}_h)\rbrace$,
$ \widehat{\upsilon}_n(h)=\widehat{\phi}^2_h \lbrace \psi^{\prime}(\widehat{y}_n(h)\widehat{\phi}_h) + \psi^{\prime}((1-\widehat{y}_n(h))\widehat{\phi}_h)\rbrace$,
and
$\psi^{\prime\prime} (\cdot)$
is the derivative
of the trigamma function.

The construction of the
BCa prediction interval
for the \barma~model is
based on the algorithm presented
in~\cite{espinheira2014}
and~\cite{espinheira2017e},
where the prediction limits are given by:
\begin{align*}
LL_h^b
&=
\frac{1}
{ 1+\exp \left\{ -\widehat{y}^\ast_{n}(h) - \delta_{(\alpha/2)} \sqrt{\widehat{\nu}_{n}(h)} \right\}}
,
\\
UL_h^b
&=
\frac{1}
{1+\exp \left\{- \widehat{y}^\ast_{n}(h) - \delta_{(1-\alpha/2)} \sqrt{\widehat{\nu}_{n}(h)} \right\}}
,
\end{align*}
where
$\tilde{\alpha}=\Phi\left\{ \widehat{z}_0 + [ (\widehat{z}_0 + z_\alpha)^{-1}
-
\widehat{a}]^{-1}\right\}$,
$\Phi(\cdot)$
is the
cumulative distribution function
of the standard normal distribution,
$\tilde{\alpha}=\Phi\left\{ \widehat{z}_0 + [ (\widehat{z}_0 + z_\alpha)^{-1}-\widehat{a}]^{-1}\right\}$,
$\delta_{ (\alpha/2)} =\mathcal{R}_{3(\tilde{\alpha}/2)}
(y_{n}^{\ast b}(h),\widehat{y}^b_n(h))$,
$\delta_{ (1-\alpha/2)} =\mathcal{R}_{3(1-\tilde{\alpha}/2)}
(y_{n}^{\ast b}(h),\widehat{y}^b_n(h))$,
The
quantity
$y_{n}^{\ast b}(h)$
is defined in~\cite{espinheira2017e},
$\widehat{y}^\ast_{n}(h) = \log(\widehat{y}_{n}(h)/(1-\widehat{y}_{n}(h)))$,
and
$\widehat{\nu}_n (h) =
\psi^{\prime}(\widehat{y}_n(h)\widehat{\phi}_h)
+
\psi^{\prime}(\left(1-\widehat{y}_n(h)\right)\widehat{\phi}_h)$.

\subsection{Bootstrap Percentile Interval}
\label{s:percentil}

Another way to construct prediction intervals
is by finding
the percentiles of the~$B$
replications in the considered experiment~\citep{Davison1997}.
For instance,
one can consider the use of
percentile intervals~\citep{Efron1994},
which are
based on~$B$ bootstrap resamplings
of the predicted
values~$\widehat{y}_n(h)$,
furnishing:
\begin{align*}
[\widehat{y}_{n}^{I}(h);\widehat{y}_{n}^{S}(h)]
,
\end{align*}
where
$\widehat{y}_{n}^{I}(h)$
and
$\widehat{y}_{n}^{S}(h)$
are
the
$\alpha / 2 $
and
$1 - \alpha / 2 $
quantiles
of
of $y_{n+h}^b$,
$b = 1, 2, \ldots , B$.
The percentile
prediction intervals
were
constructed considering two methods
of bootstrap resampling.
One approach is based
on block
resampling~\citep{Davison1997},
and
the other one is the
residual method with
bootstrap samples
computed as
\begin{align*}
y_{t}^b=  g ^{-1} \left( \widehat{y}_t^\ast +
\mathcal{R}_1^\ast (y_{t},\widehat{\mu}_t) \right),
\end{align*}
where the residuals
$\mathcal{R}_1^\ast
(y_{t},\widehat{\mu}_t)$
are uniformly
random
draws
from
$\mathcal{R}_1 (y_1,\widehat{\mu}_1),
\mathcal{R}_1 (y_2,\widehat{\mu}_2),
\ldots,
\mathcal{R}_1 (y_n,\widehat{\mu}_n)$
and
$\widehat{y}_t^\ast=\log\{\widehat{\mu}_t/(1-\widehat{\mu}_t)$.
Bootstrap forecast realizations for future values~$y_{n+h}^b$
are given by:
\begin{align}
\label{e:futurosboot}
\widehat{y}_{n}^b(h)
=
g^{-1}
\left(
\widehat{\beta}^b
+
\sum_{i=1}^{p}\widehat{\varphi}_i^b
g^\star(y_{n+h-i}^b)
+
\sum_{j=1}^{q}\widehat{\theta}_j^b
r_{n+h-j}^{\star b}
\right)
,
\end{align}
where
\begin{align*}
g^\star(y_{n+h-i}^b)
& =
\begin{cases}
g(\widehat{y}_{n}^b(h-i)),&\text{if $i< h$},\\
g(y_{n+h-i}^b), & \text{if $i\geq h$},
\end{cases}
\end{align*}
and
\begin{align*}
r_{n+h-j}^{\star b}
&=
\begin{cases}
g(y_{n+h-j}^b) - g(\widehat{y}_n^b(h-j)),&\text{if $j< h$},\\
g(y_{n+h-j}^b) - g(\widehat{\mu}^b_{n+h-j}), & \text{if $j\geq h$}.
\end{cases}
\end{align*}

\section{
	Prediction Intervals Comparison}
\label{s:num}

The
numerical
evaluation of the
\barma~model
prediction intervals
was performed
considering
Monte Carlo simulations.
The number~$R$ of Monte Carlo replications
and the number of bootstrap
resampling
were both set equal to $1{,}000$,
as suggested
in~\cite{Hotta2016}.
The adopted
sample size
forecast horizon,
and
significance level
were~$n = 100$,~$H=10$,
and~$\alpha = 0.10$,
respectively.
The computational implementation was developed
in~\texttt {R}
language~\citep{R2012}.

We generated~$n+H$
data points,
where the ending~$H$ observations
were
used to assess
the prediction intervals.
The structure of the mean
of the \barma~model was given according to
Eq.~\eqref{E:modelo},
employing the logit link function
$\operatorname{logit}(\mu_t)=\log(\frac{\mu_t}{1-\mu_t})$.
We analyzed the following simulation scenarios:
\begin{enumerate}[I.]
	\item \label{i:m3} \barma$(1,1)$: $\beta=-0.3$, $\varphi_1=-0.4$,
	and
	$\theta_1=0.3$;
	\item \label{i:m4} \barma$(1,1)$: $\beta=0.95$, $\varphi_1=0.65$,
	and
	$\theta_1=-0.95$;
	\item \label{i:m1} $\beta$AR$(2)$: $\beta=-0.3$, $\varphi_1= 0.8$,
	and
	$\varphi_2=-0.8$;
	\item \label{i:m2} $\beta$AR$(2)$: $\beta=0.9$, $\varphi_1=0.3$,
	and
	$\varphi_2=0.3$;
	\item \label{i:m5} $\beta$MA$(2)$: $\beta=-0.8$, $\theta_1=0.8$,
	and
	$\theta_2=-0.8$;
	\item \label{i:m6} $\beta$MA$(2)$: $\beta=1.5$, $\theta_1=-0.2$,
	and
	$\theta_2=0.6$.
\end{enumerate}

The above
choice of parameters
aims at capturing
different characterizations
of~$\mu$.
For example,
Scenarios~\ref{i:m3},~\ref{i:m1}, and~\ref{i:m5}
adopt
$\mu \approx 0.4$
(almost symmetric distribution),
and
Scenarios~\ref{i:m4},~\ref{i:m2}, and~\ref{i:m6}
employ $\mu \approx 0.9$~(asymmetric distribution).
All models adopt~$\phi=120$.
We also experimented with
other values of
$\phi$, $\alpha$, and $n$;
however the results were
not significantly different
and are omitted for brevity.
The
results
for
Scenarios~\ref{i:m3} and~\ref{i:m4}
were separated
as representative cases
for the parameter range of the
\barma~model;
results for remaining scenarios
are detailed in the Appendix.

To evaluate the prediction intervals,
coverage rates $(\text{CR}_h)$
for each interval at coverage level of $90\%$
were computed
according to:
\begin{align*}
\text{CR}_h = \frac{\# (LL_h < y_{n+h} < UL_h) }{R}
.
\end{align*}
It is desirable that
$\text{CR}_h$ approaches
the nominal coverage level $(1-\alpha)=0.90$.
Additionally,
we computed
the rates
for
the actual future value
above the upper limit
or
below the lower limit.
Thus,
we have the
average upper interval $(\text{CR}_h^U)$
and
the average lower interval $(\text{CR}_h^L)$
furnished
by~\citep{Hotta2016, pascual2004}:
\begin{align*}
\text{CR}_h^U & = \frac{\# (UL_h < y_{n+h}) }{R}, \\
\text{CR}_h^L & = \frac{\# (LL_h > y_{n+h}) }{R}.
\end{align*}
Good prediction intervals,
exhibit:
$\text{CR}_h^U \approx \text{CR}_h^L \approx \alpha/2$.
The average length $(\text{A}_h)$ of the intervals
was also computed,
according to~\citep{Hotta2016}:
\begin{align*}
\text{A}_h
&=
\frac{1}{R}
\sum \limits^R _ {i=1} (UL_h^{(i)} - LL_h^{(i)})
.
\end{align*}
The quantity $\text{A}_h$ is
expected
to be as small as possible.
The
measures
$\text{CR}_h$,
$\text{CR}_h^U$,
$\text{CR}_h^L$,
and
$\text{A}_h$
were also considered
in~\cite{Thombs1990},~\cite{pascual2004},~\cite{Pascual2006},
and~\cite{Hotta2016}
for evaluation
of the prediction intervals.

To obtain an overall evaluation
performance of the predicted intervals,
we considered
the prediction interval coverage probability~(PICP)
as a figure of merit,
which is given by~\citep{quan2014}:
\begin{align*}
\text{PICP} =
\frac{
	1 }{H}
\sum \limits_{h=1}^H
\text{CR}_h
.
\end{align*}
PICP values that are close
to the nominal coverage level
indicate good
performance.
Finally,
we
computed
the
prediction interval normalized average width~(PINAW)
measure,
which is defined as~\citep{quan2014}:
\begin{align*}
\text{PINAW} = \frac{
	\frac{1}{H}
	\sum \limits_{h=1}^H
	\text{A}_h}{\text{max} (y_t) - \text{min} (y_t)}
.
\end{align*}
The quantity PINAW
is sought to be as
small as possible.

Table~\ref{t:barma04} displays
the Monte Carlo simulation results for Scenario~\ref{i:m3}.
This model consists of
one autoregressive term
and
one moving average term
with $\mu \approx 0.4$
(almost symmetric distribution).
The BCa
prediction
interval
shows
values of coverage rate closer to the nominal values
when compared to other prediction intervals.
The
Qbeta, BCa,
and
residual percentile prediction intervals exhibit
$\text{CR}_h^L$ closer to $\text{CR}_h^U$
when compared with the remaining intervals.

\begin{table}

	\caption{Estimated coverage rates, average length, and balanced of
		the prediction intervals for the Scenario~\ref{i:m3}
	}
	\tablesize
	\centering
	\label{t:barma04}
	\begin{tabular}{lcccccccccccc}
		\toprule
		$h$ & $1$ & $2$ & $3$ & $4$ & $5$ & $6$ & $7$ & $8$ & $9$ & $10$ \\
		\midrule
		\multicolumn{11}{c}{BJ Prediction Interval}\\
		\midrule
		$\text{CR}_h$	&	$0.897$ & $0.879$ & $0.882$ & $0.885$ & $0.863$ & $0.869$ & $0.878$ & $0.883$ & $0.863$ & $0.884$\\
		$\text{A}_h$	&	$0.150$ & $0.177$ & $0.193$ & $0.207$ & $0.215$ & $0.223$ & $0.228$ & $0.233$ & $0.236$ & $0.240$ \\
		$\text{CR}_h^L$	&	$0.046$ & $0.061$ & $0.053$ & $0.064$ & $0.069$ & $0.070$ & $0.065$ & $0.062$ & $0.072$ & $0.057$ \\
		$\text{CR}_h^U$ & $0.057$ & $0.060$ & $0.065$ & $0.051$ & $0.068$ & $0.061$ & $0.057$ & $0.055$ & $0.065$ & $0.059$	\\
		\midrule
		\multicolumn{11}{c}{Qbeta Prediction Interval}\\
		\midrule
		$\text{CR}_h$	&	$0.890$ & $0.874$ & $0.874$ & $0.882$ & $0.860$ & $0.861$ & $0.867$ & $0.879$ & $0.855$ & $0.873$	\\
		$\text{A}_h$	&	$0.147$ & $0.173$ & $0.189$ & $0.202$ & $0.211$ & $0.218$ & $0.223$ & $0.228$ & $0.231$ & $0.234$	\\
		$\text{CR}_h^L$	&	$0.050$ & $0.064$ & $0.057$ & $0.063$ & $0.070$ & $0.072$ & $0.067$ & $0.062$ & $0.073$ & $0.058$	\\
		$\text{CR}_h^U$	& $0.060$ & $0.062$ & $0.069$ & $0.055$ & $0.070$ & $0.067$ & $0.066$ & $0.059$ & $0.072$ & $0.069$		\\
		\midrule
		\multicolumn{11}{c}{BPE Prediction Interval}\\
		\midrule
		$\text{CR}_h$	&	$0.853$ & $0.806$ & $0.792$ & $0.760$ & $0.731$ & $0.746$ & $0.721$ & $0.740$ & $0.711$ & $0.700$	\\
		$\text{A}_h$	&	$0.149$ & $0.170$ & $0.178$ & $0.190$ & $0.193$ & $0.200$ & $0.201$ & $0.207$ & $0.207$ & $0.211$	\\
		$\text{CR}_h^L$	&	$0.078$ & $0.096$ & $0.110$ & $0.113$ & $0.152$ & $0.127$ & $0.156$ & $0.123$ & $0.164$ & $0.138$	\\
		$\text{CR}_h^U$	&	$0.069$ & $0.098$ & $0.098$ & $0.127$ & $0.117$ & $0.127$ & $0.123$ & $0.137$ & $0.125$ & $0.162$	\\
		\midrule
		\multicolumn{11}{c}{BCa Prediction Interval}\\
		\midrule
		$\text{CR}_h$	&	$0.904$ & $0.914$ & $0.913$ & $0.906$ & $0.898$ & $0.904$ & $0.918$ & $0.917$ & $0.899$ & $0.908$	\\
		$\text{A}_h$	&	$0.172$ & $0.203$ & $0.223$ & $0.238$ & $0.248$ & $0.256$ & $0.262$ & $0.268$ & $0.272$ & $0.275$	\\
		$\text{CR}_h^L$	&	$0.038$ & $0.050$ & $0.041$ & $0.052$ & $0.050$ & $0.048$ & $0.046$ & $0.036$ & $0.054$ & $0.043$	\\
		$\text{CR}_h^U$	&	$0.058$ & $0.036$ & $0.046$ & $0.042$ & $0.052$ & $0.048$ & $0.036$ & $0.047$ & $0.047$ & $0.049$	\\
		\midrule
		\multicolumn{11}{c}{Block Percentile Prediction Interval}\\
		\midrule
		$\text{CR}_h$	&	$0.884$& $0.879$ & $0.878$ & $0.874$ & $0.862$ & $0.861$ & $0.863$ & $0.863$ & $0.869$ & $0.877$	\\
		$\text{A}_h$	&   $0.251$ & $0.252$ & $0.251$ & $0.251$ & $0.250$ & $0.250$ & $0.249$ & $0.249$ & $0.249$ & $0.249$	\\
		$\text{CR}_h^L$	&	$0.055$ & $0.062$ & $0.069$ & $0.071$ & $0.069$ & $0.074$ & $0.076$ & $0.063$ & $0.073$ & $0.055$	\\
		$\text{CR}_h^U$	&  $0.061$ & $0.059$ & $0.053$ & $0.055$ & $0.069$ & $0.065$ & $0.061$ & $0.074$ & $0.058$ & $0.068$		\\
		\midrule
		\multicolumn{11}{c}{Residual Percentile Prediction Interval}\\
		\midrule
		$\text{CR}_h$	&	$0.951$ & $0.931$ & $0.920$ & $0.904$ & $0.890$ & $0.883$ & $0.889$ & $0.883$ & $0.874$ & $0.890$	\\
		$\text{A}_h$	&		$0.226$ & $0.231$ & $0.235$ & $0.237$ & $0.240$ & $0.242$ & $0.243$ & $0.244$ & $0.246$ & $0.246$ \\
		$\text{CR}_h^L$	&	$0.020$ & $0.040$ & $0.035$ & $0.053$ & $0.053$ & $0.061$ & $0.060$ & $0.055$ & $0.066$ & $0.053$	\\
		$\text{CR}_h^U$	&	$0.029$ & $0.029$ & $0.045$ & $0.043$ & $0.057$ & $0.056$ & $0.051$ & $0.062$ & $0.060$ & $0.057$	\\
		\bottomrule
	\end{tabular}
\end{table}

\begin{figure}
	\centering
	\subfigure[Comparison of the $\text{CR}_h$]
	{\label{f:perbar3}\includegraphics[width=0.5\textwidth]{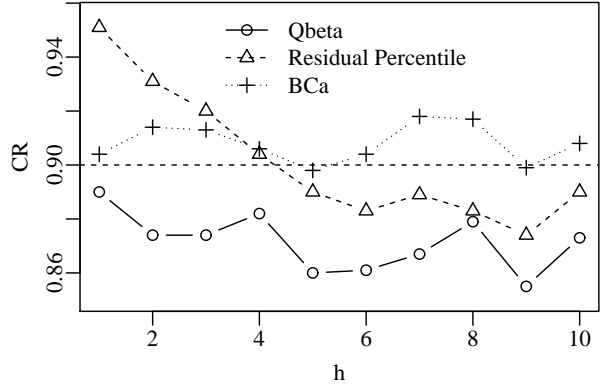}}
	\subfigure[Comparison of the $\text{A}_h$]
	{\label{f:perbar_h3}\includegraphics[width=0.5\textwidth]{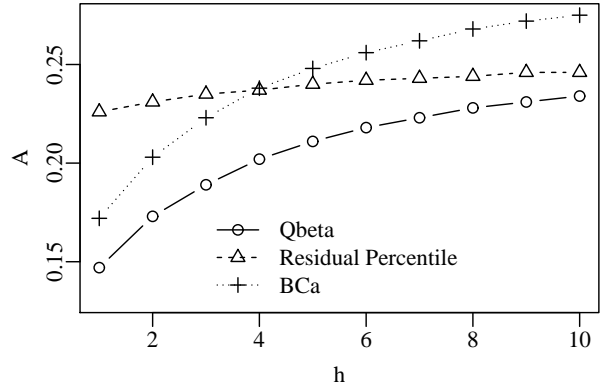}} \\
	\caption{Comparison of  Qbeta,
		Residual Percentile
		and BCa
		prediction intervals
		for the Scenario~\ref{i:m3}.}\label{f:barma04}
\end{figure}

Figure~\ref{f:barma04} features a graphical summary of the considered
measures for Qbeta, residual percentile, and BCa prediction intervals.
These three intervals were separated
because they
behave
similarly
and
their associated
coverage rates $\text{CR}_h$
are
closer to the nominal values
for both assessed scenarios.
The coverage rates
are shown in Figure~\ref{f:perbar3},
being close to the nominal value of $0.90$.
The BCa prediction interval
presents
$\text{CR}_h$
values
closer to the nominal
level in~$90\%$
of the occurrences
when compared to
the Qbeta and residual percentile prediction intervals.
Additionally,
the Qbeta and residual percentile
prediction intervals
present the largest variations
in terms of~$\text{CR}_h$.
Figure~\ref{f:perbar_h3}
presents the obtained average lengths;
all the
considered
prediction intervals
present
average length
values close to zero.
Among
the
evaluated
prediction intervals,
the Qbeta prediction interval
offers
smaller values
of average length
but
shows
coverage rates farther
from
the nominal value, $0.90$.

Table~\ref{t:barma09}
brings
the Monte Carlo simulation results for Scenario~\ref{i:m4},
which
possesses
one autoregressive term
and
one moving average term
for
$\mu \approx 0.9$ (asymmetric distribution).
The prediction intervals
offer
similar results,
except for the BJ prediction interval.
The $\text{CR}_h$ values for the BJ prediction interval
have noticeably departed from the values of coverage level
obtained from the other prediction intervals.
For instance,
at $h=6$ and $h=9$,
the $\text{CR}_h$ values for the BJ
prediction interval
are
equal to $0.385$ and $0.384$, respectively.
A possible reason for the distortion of the BJ
prediction interval
is the fact that the distribution of~$y$ presents a greater asymmetry,
due to the fact that $\mu=0.9$ is closer
to the upper limit of the unit interval.
Therefore,
the normality assumption becomes inappropriate.
As a consequence,
BJ
prediction interval
performs
poorly
when $\mu$ is close to $0$ or $1$.
For the EPB and residual percentile prediction intervals,
the values of $\text{CR}_h^L$ and $\text{CR}_h^U$
are similar
when compared with
the remaining prediction intervals.

\begin{table}
	\centering
	\caption{Estimated coverage rates, average length, and balanced
		of the prediction intervals for the Scenario~\ref{i:m4} }
	\tablesize
	\label{t:barma09}
	\begin{tabular}{lcccccccccccc}
		\toprule
		$h$ & $1$ & $2$ & $3$ & $4$ & $5$ & $6$ & $7$ & $8$ & $9$ & $10$ \\
		\midrule
		\multicolumn{11}{c}{BJ Prediction Interval}\\
		\midrule
		$\text{CR}_h$	&	$0.395$ & $0.423$ & $0.438$ & $0.397$ & $0.447$ & $0.385$ & $0.430$ & $0.435$ & $0.384$ & $0.415$	\\
		$\text{A}_h$	&	$0.026$ & $0.027$ & $0.026$ & $0.026$ & $0.026$ & $0.026$ & $0.026$ & $0.026$ & $0.026$ & $0.026$\\
		$\text{CR}_h^L$	&	$0.234$ & $0.252$ & $0.272$ & $0.304$ & $0.293$ & $0.315$ & $0.311$ & $0.288$ & $0.325$ & $0.296$ 	\\
		$\text{CR}_h^U$	&	$0.371$ & $0.325$ & $0.290$ & $0.299$ & $0.260$ & $0.300$ & $0.259$ & $0.277$ & $0.291$ & $0.289$ \\
		\midrule
		\multicolumn{11}{c}{Qbeta Prediction Interval}\\
		\midrule
		$\text{CR}_h$	&	$0.924$ & $0.926$ & $0.936$ & $0.935$ & $0.932$ & $0.925$ & $0.935$ & $0.941$ & $0.934$ & $0.931$	\\
		$\text{A}_h$	&	$0.082$ & $0.086$ & $0.088$ & $0.088$ & $0.088$ & $0.088$ & $0.088$ & $0.088$ & $0.088$ & $0.088$	\\
		$\text{CR}_h^L$	&	$0.027$ & $0.041$ & $0.043$ & $0.038$ & $0.047$ & $0.044$ & $0.052$ & $0.043$ & $0.042$ & $0.046$	\\
		$\text{CR}_h^U$	&	$0.049$ & $0.033$ & $0.021$ & $0.027$ & $0.021$ & $0.031$ & $0.013$ & $0.016$ & $0.024$ & $0.023$	\\
		\midrule
		\multicolumn{11}{c}{BPE Prediction Interval}\\
		\midrule
		$\text{CR}_h$	&	$0.843$ & $0.847$ & $0.886$ & $0.864$ & $0.864$ & $0.861$ & $0.857$ & $0.882$ & $0.848$ & $0.862$	\\
		$\text{A}_h$	&	$0.101$ & $0.078$ & $0.081$ & $0.075$ & $0.076$ & $0.074$ & $0.076$ & $0.074$ & $0.074$ & $0.074$	\\
		$\text{CR}_h^L$	&	$0.005$ & $0.049$ & $0.056$ & $0.074$ & $0.085$ & $0.086$ & $0.100$ & $0.085$ & $0.106$ & $0.097$	\\
		$\text{CR}_h^U$	&	$0.152$ & $0.104$ & $0.058$ & $0.062$ & $0.051$ & $0.053$ & $0.043$ & $0.033$ & $0.046$ & $0.041$	\\
		\midrule
		\multicolumn{11}{c}{BCa Prediction Interval}\\
		\midrule
		$\text{CR}_h$	&	$0.897$ & $0.903$ & $0.895$ & $0.894$ & $0.900$ & $0.882$ & $0.899$ & $0.913$ & $0.897$ & $0.895$	\\
		$\text{A}_h$	&	$0.078$ & $0.078$ & $0.080$ & $0.078$ & $0.080$ & $0.078$ & $0.079$ & $0.079$ & $0.078$ & $0.078$	\\
		$\text{CR}_h^L$	& $0.073$ & $0.068$ & $0.081$ & $0.067$ & $0.071$ & $0.067$ & $0.075$ & $0.061$ & $0.062$ & $0.066$	\\
		$\text{CR}_h^U$	&	$0.030$ & $0.029$ & $0.024$ & $0.039$ & $0.029$ & $0.051$ & $0.026$ & $0.026$ & $0.041$ & $0.039$	\\
		\midrule
		\multicolumn{11}{c}{Block Percentile Prediction Interval}\\
		\midrule
		$\text{CR}_h$	&	$0.950$ & $0.926$ & $0.945$ & $0.924$ & $0.924$ & $0.915$ & $0.932$ & $0.933$ & $0.929$ & $0.926$	\\
		$\text{A}_h$	&	$0.096$ & $0.090$ & $0.090$ & $0.088$ & $0.088$ & $0.088$ & $0.088$ & $0.088$ & $0.088$ & $0.088$	\\
		$\text{CR}_h^L$	&	$0.015$ & $0.035$ & $0.021$ & $0.029$ & $0.032$ & $0.029$ & $0.034$ & $0.032$ & $0.030$ & $0.029$	\\
		$\text{CR}_h^U$	& $0.035$ & $0.039$ & $0.034$ & $0.047$ & $0.044$ & $0.056$ & $0.034$ & $0.035$ & $0.041$ & $0.045$		\\
		\midrule
		\multicolumn{11}{c}{Residual Percentile Prediction Interval}\\
		\midrule
		$\text{CR}_h$	&	$0.940$ & $0.915$ & $0.916$ & $0.902$ & $0.906$ & $0.887$ & $0.906$ & $0.908$ & $0.899$ & $0.899$	\\
		$\text{A}_h$	&	$0.088$ & $0.084$ & $0.082$ & $0.080$ & $0.080$ & $0.080$ & $0.080$ & $0.080$ & $0.080$ & $0.080$	\\
		$\text{CR}_h^L$	&	$0.029$ & $0.043$ & $0.050$ & $0.044$ & $0.054$ & $0.048$ & $0.060$ & $0.051$ & $0.048$ & $0.051$	\\
		$\text{CR}_h^U$	&	$0.031$ & $0.042$ & $0.034$ & $0.054$ & $0.040$ & $0.065$ & $0.034$ & $0.041$ & $0.053$ & $0.050$	\\
		\bottomrule
	\end{tabular}
\end{table}

\begin{figure}
	\centering
	\subfigure[Comparison of the $\text{CR}_h$]
	{\label{f:perbar4}\includegraphics[width=0.5\textwidth]{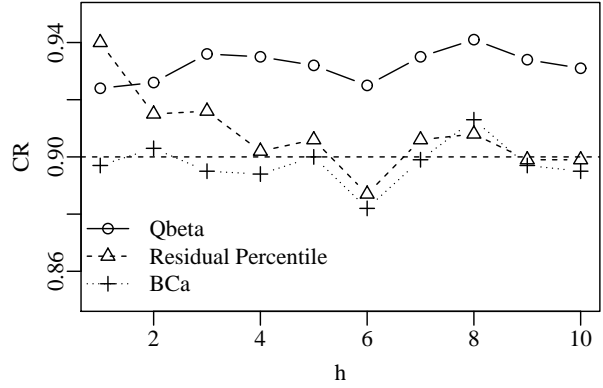}}
	\subfigure[Comparison of the $\text{A}_h$]
	{\label{f:perbar_h4}\includegraphics[width=0.5\textwidth]{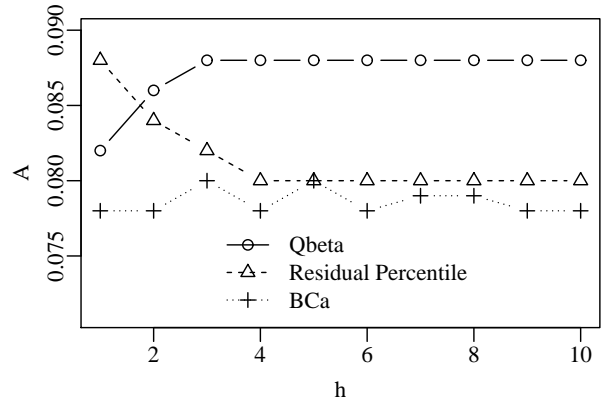}} \\
	\caption{Comparison of  Qbeta,
		Residual Percentile
		and BCa
		prediction intervals
		for the Scenario~\ref{i:m4}.}\label{f:barma09}
\end{figure}

Figure~\ref{f:perbar4}
shows that
the
residual percentile
and
BCa
prediction intervals
return
values for coverage rates closer to the nominal values.
As discussed in Scenario I,
the BCa prediction interval
presents~$\text{CR}_h$
values
closer to the nominal
level
in comparison with
the other two considered prediction intervals
($60\%$ of the observations)
and the smallest variation in the~$\text{CR}_h$.
Additionally,
Figure~\ref{f:perbar_h4} presents
the average length results;
the BCa prediction interval
is linked
to the
lowest average length.
Thus,
the BCa prediction interval
offers
good values of $\text{CR}_h$
and
a
small average length.

Finally, Table~\ref{t:comp}
presents the measurements
of
PICP and PINAW.
The  evaluated figures of merit
are in agreement with the
results based on the coverage rate and average length.
The BCa and
residual percentile
prediction intervals
outperformed
the remaining methods
for both considered scenarios.
Moreover,
a combined analysis of
PICP and PINAW
confirms the above conclusions,
i.e.,
PICP  is close to the nominal level
and PINAW presenting small values.

\begin{table}
	\centering
	\caption{
		Numerical results of the figures of merit
		PICP and PINAW
		used to evaluate the prediction intervals}
	\tablesize
	\label{t:comp}
	\begin{tabular}{ccccccc}
		\toprule
		& \multicolumn{4}{c}{Figure of merit} \\
		\midrule
		Prediction Interval &
		PICP & PINAW &
		PICP & PINAW &
		\\
		\midrule
		& \multicolumn{2}{c}{Scenario I} &
		\multicolumn{2}{c}{Scenario II} \\
		\midrule
		BJ & $0.8783$  & $0.2102$ &
		$0.4149$ & $0.0261$
		\\
		Qbeta & $0.8715$ & $0.2056$ &
		$0.9319$& $0.0872$
		\\
		BPE & $0.7560$ &$0.1906$ &
		$0.8614$ &$0.0783$
		\\
		BCa & $0.9081$ &$0.2417$  &
		$0.8975$ &$0.0786$
		\\
		Block Percentile & $0.8710$  & $0.2501$  &
		$0.9304$ & $0.0892$
		\\
		Residual Percentile & $0.9015$ & $0.2390$ &
		$0.9078$ & $0.0814$
		\\

		\bottomrule
	\end{tabular}
\end{table}

In a general manner,
the
BJ prediction interval
presents
values of coverage rates
farther from nominal values
when considering asymmetric
distributions ($\mu \approx 0.9 $).
Among the
non-bootstrap-based
intervals,
the Qbeta
prediction interval
presents
smaller distortions
when
compared to the interval BJ.
On the other hand,
bootstrap prediction intervals
show
values of $\text{CR}_h$
closer to
the coverage level values
in comparison
with
the prediction intervals
without bootstrapping.
Under Scenarios~\ref{i:m1} and~\ref{i:m2},
which consider only autoregressive terms,
BPE prediction interval
generates
values of $\text{CR}_h$
distant from the nominal values
when compared
to the results from
the models that consider moving average terms.
In the Appendix, we provide detailed results.
This way,
in the \barma~model,
the use of
moving average terms
generates
more accurate results
in
BPE prediction interval.

The discussed methods
presented variable performances
depending on the selected scenario.
However,
BCa prediction interval
in the considered
models
shows smaller average length,
presents~$\text{CR}_h^L$ close to $\text{CR}_h^U$,
and mainly
with values of $\text{CR}_h$ constant
and close to the coverage level values of the intervals.
In terms of
$\text{CR}_h^L$, $\text{CR}_h^U$,
and
$\text{CR}_h$,
the residual prediction interval
shows performance similarly
to the BCa prediction interval.
However,
when average length is considered,
the BCa prediction interval maintains its
good performance;
whereas the
residual prediction interval offers
poorer results.
Thus,
BCa prediction interval
generally presents
the best performance under all the evaluated scenarios.
As a consequence,
we identify the
BCa prediction interval
as a method capable of
performing well according
to all discussed figures of merit.

\section{Water Level Prediction}
\label{s:aplicacao}

This section presents an application of
the proposed prediction intervals
to measured
data.
The employed data set corresponds
to the water level
of the reservoirs
from the Metropolitan Area
of S\~ao Paulo, Brazil (Cantareira System)~\citep{sabesp}.
The
water level
is the proportion of water available in
relation to the total storage
capacity of the reservoir.
The
data were
measured
in the period of
January 2003 to
July 2015,
consisting of
151~monthly observations.
The
ending
ten
observations were separated for
assessment of the considered prediction intervals.

\begin{figure}
	\centering
	\subfigure[Original series]
	{\label{f:serie2}\includegraphics[width=0.47\textwidth]{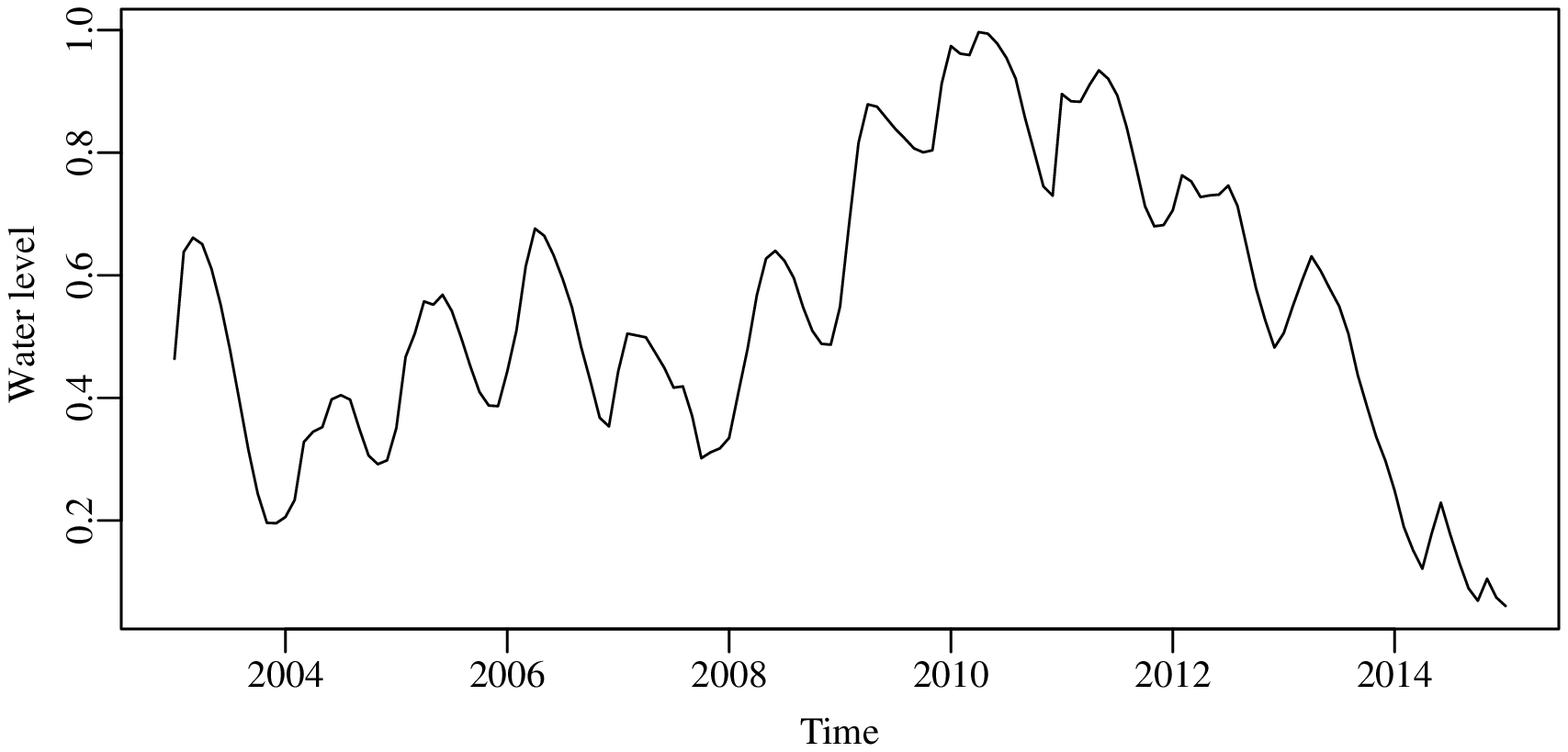}}
	\subfigure[Histogram]
	{\label{f:hist2}\includegraphics[width=0.47\textwidth]{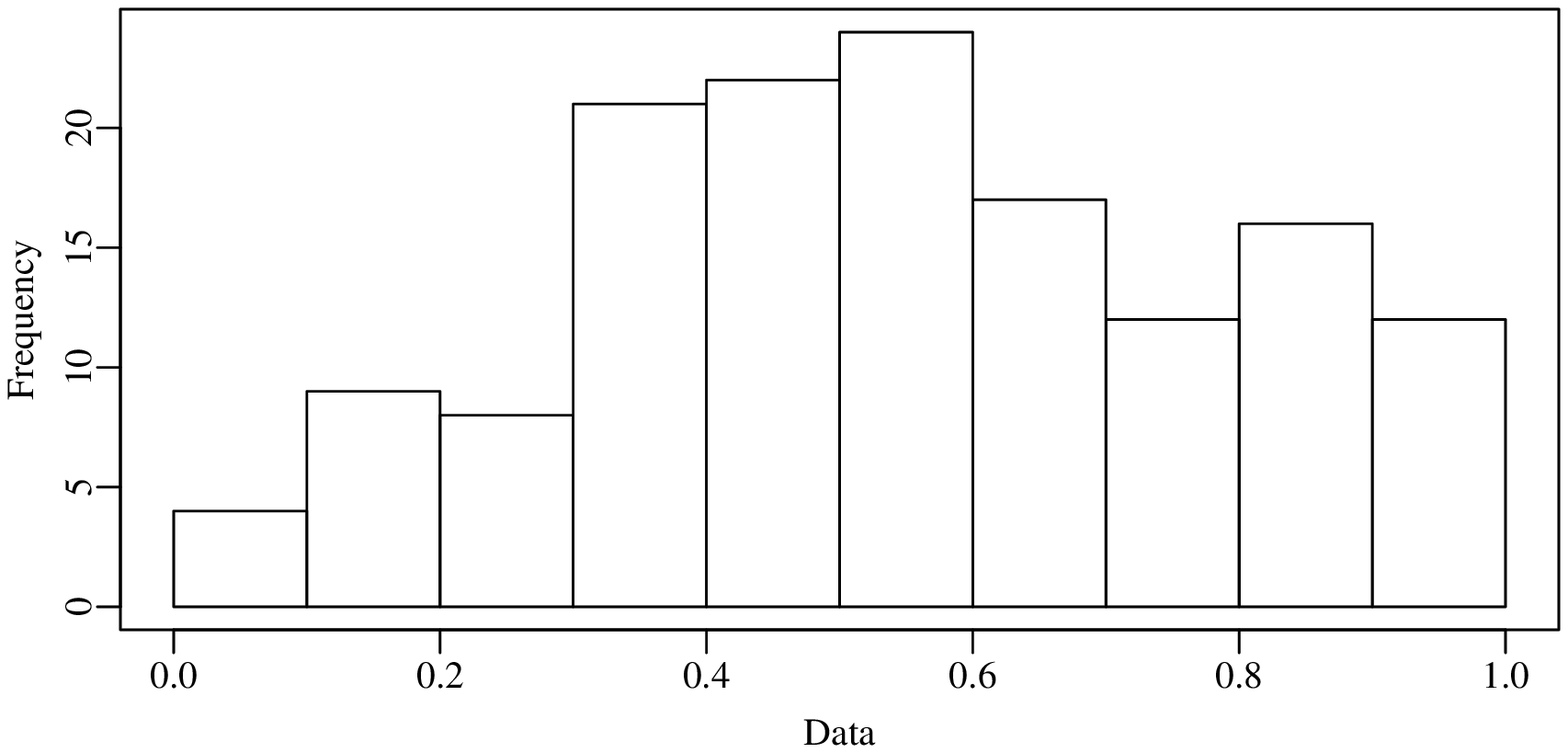}} \\
	\subfigure[Sampling ACF]
	{\label{f:fac2}\includegraphics[width=0.47\textwidth]{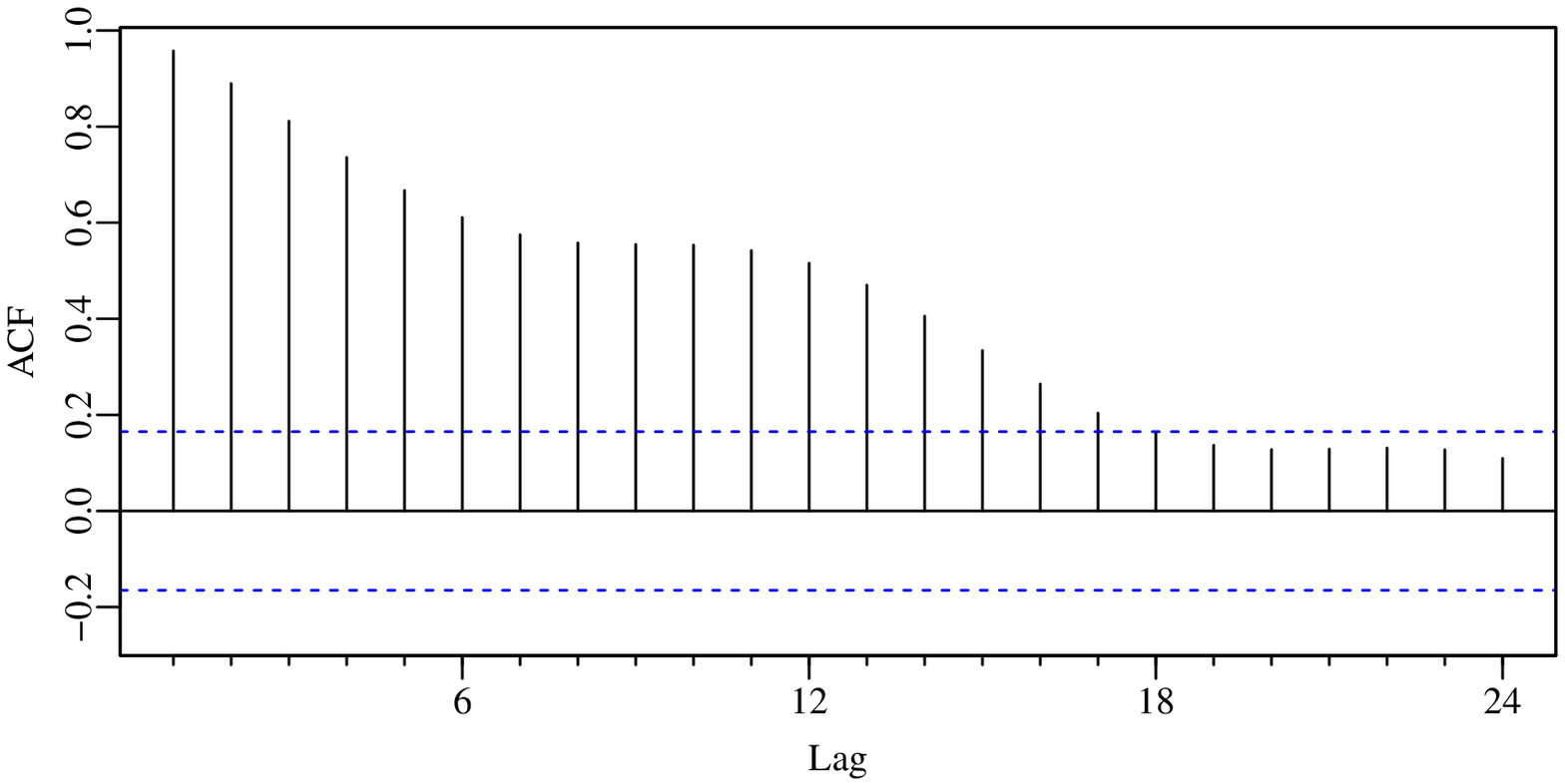}}
	\subfigure[Sampling PACF]
	{\label{f:facp2}\includegraphics[width=0.47\textwidth]{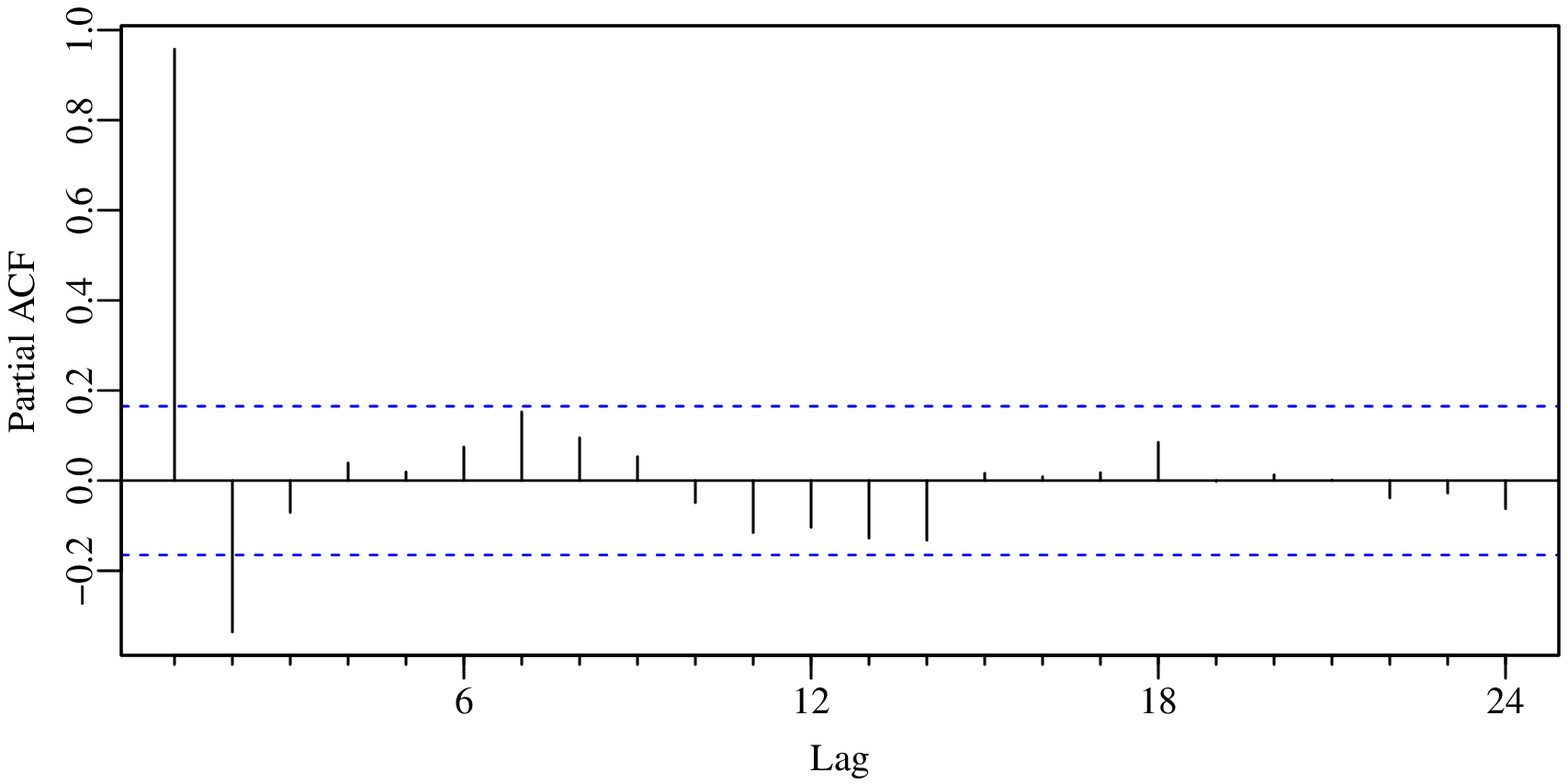}}
	\caption{Line chart, histogram, and correlograms of the
		time series
		of the water levels of the Cantareira System.}
	\label{f:dados2}
\end{figure}

Figure~\ref{f:serie2} displays
the considered
time series
with
unconditional mean
of~$0.530$.
The kurtosis
and skewness
are equal to $-0.858$
and $0.054$, respectively,
indicating that the unconditional
distribution of the
data
has shorter
tails.
This fact can
also be
verified
on the
data
histogram in Figure~\ref{f:hist2}.
Therefore,
selecting
a model
or a prediction interval
that
assumes normality would not be suitable,
leading to less reliable conclusions.
The maximum value
of the
data
was~$0.997$,
observed in April~2010,
and the lower volume,~$0.061$, in January~2015.
Figures~\ref{f:fac2} and~\ref{f:facp2}
present
the sampling autocorrelation function (ACF)
and the sampling partial autocorrelation function (PACF),
respectively.

The model selection
was based on
the three-stage iterative Box-Jenkins methodology~\citep{Box2008},
i.e., identification (considering an exhaustive search aiming at minimizing the~AIC),
estimation, and diagnostic checking.
Adopting a
significance level of~$p=0.1$
and
restringing
the search space
to models with orders less or equal to~$12$,
we successfully adjusted the employed data using the
autoregressive terms~$g(y_{t-1})$,~$g(y_{t-2})$,~$g(y_{t-3})$,
and~$g(y_{t-4})$, and considering the logit link function.
The diagnostic analysis of the
fitted
model
was
based on the standardized residual~$\mathcal{R}_{2}(y_t,\widehat{\mu}_t)$.
If the model is correct,
then the residual is approximately
normal
with unit variance and around zero.
Table~\ref{T:ajuste12} presents
the fit of the selected model and the diagnostic analysis.
Considering the
Lagrange Multiplier,
Box-Pierce, and
Ljung-Box tests,
the residuals of the fitted model do not exhibit
conditional heteroscedasticity or autocorrelation.
To perform
the diagnostic tests,
we
followed the methodology
proposed in~\cite{hyndman2018forecasting}
to define the number of lags,
which is given by:
$
\text{Number of lags}
=
\min
\left(
10, n/5
\right)
$.
As the employed data set
consists of~$151$
monthly observations,
the number of lags is equal to~$10$.

\begin{table}
	\centering
	\tablesize
	\caption{
		\barma~model adjusted for the water levels of the Cantareira System.}
	\label{T:ajuste12}
	\begin{tabular}{lcccccc}
		\toprule
		 &${\alpha}$ & ${\varphi}_1$ &  ${\varphi}_{2}$   & ${\varphi}_{3}$   &
		${\varphi}_{4}$ &
		${\phi}$\\
		\midrule
		 Estimator &      $0.0077 $ &   $1.3838$ &  $ -0.6626$ &   $ 0.4940$ &   $-0.2706$ &   $86.4070$        \\
	     Standard Error & $0.0182$ & $ 0.0130$ &  $0.0219$ &  $0.0307$ &  $0.0282$ & $10.4190$  \\
		\midrule
		\multicolumn{7}{c}{Diagnostic analysis}\\
		\midrule
		\multicolumn{2}{c}{Test} && &  &&$p$-value \\
		\midrule
		\multicolumn{2}{c}{Lagrange Multiplier} &  &&& & $0.9624$ \\
		\multicolumn{2}{c}{Box-Pierce} & & & &&  $0.1341$ \\
		\multicolumn{2}{c}{Ljung-Box} &  && && $0.1153$\\
		\bottomrule
	\end{tabular}
\end{table}

To assess the overall performance
of the evaluated predicted intervals
for water level data,
we considered
three
figures of merit,
namely:
(i)~coverage width-based criterion~(CWC)~\citep{quan2014};
(ii)~Winkler score (Score)~\citep{quan2014,winkler1972};
and
(iii)~accumulated width deviation~(AWD)~\citep{wang2018},
which are defined, respectively, as:
\begin{align*}
\text{CWC} &=
\text{PINAW} + \Phi (\text{PICP})
\exp
\left\lbrace
- \kappa [\text{PICP} -( 1- \alpha)]
\right\rbrace
,
\\
\overline{S} &=
\frac{1}{H}
\sum \limits_{h=1}^H
\mid
\text{S}_h
\mid
,
\\
\text{AWD} & =
\frac{1}{H} \sum \limits_{h=1}^H
\text{AWD}_h
,
\end{align*}
where~$\kappa$ is a
value which determines how much
penalty is assigned to prediction intervals with
a low coverage probability~\citep{khosravi2010};
$\Phi(\text{PICP}) = 0$,
for~$\text{PICP} \geq 1 - \alpha$,
and $\Phi(\text{PICP}) = 1$,
otherwise.
Additionally,~$\text{S}_h$
and~$\text{AWD}_h$
are defined, respectively, as
\begin{align*}
\text{S}_h
&=
\begin{cases}
-2 \alpha A_h - 4( LL_h - y_{n+h}),
&\text{if $y_{n+h}< LL_h$},\\
-2 \alpha A_h ,
& \text{if $LL_h <  y_{n+h}< UL_h$}, \\
-2 \alpha A_h - 4( y_{n+h} -  UL_h ),
&\text{if $ UL_h > y_{n+h}$},
\end{cases}
\end{align*}
and
\begin{align*}
\text{AWD}_h
&=
\begin{cases}
\frac{LL_h - y_{n+h}}{A_h},
&\text{if $y_{n+h}< LL_h$},\\
0 ,
& \text{if $LL_h <  y_{n+h}< UL_h$}, \\
\frac{y_{n+h} - UL_h}{A_h},
&\text{if $ UL_h > y_{n+h}$}.
\end{cases}
\end{align*}
The above measures
are expected to be as close to zero as
possible.

Table~\ref{T:pi_medidas} presents
the measured values
of PICP, PINAW, CWC, Score, and AWD
of the proposed prediction intervals.
We set~$\kappa=2$ and~$\kappa=10$
for  the CWC measure
aiming at
evaluating its behavior variations.
The Qbeta, BCa,
and  residual percentile
prediction intervals
excel in term of the
considered figures of merit.
Thus,
the prediction intervals
for
measured
water level
show
performance
similar
to the Monte Carlo simulations.
Therefore,
we recommend the use of the BCa prediction interval
in order to obtain
accurate prediction intervals.

\begin{table}
	\centering
	\tablesize
	\caption{
		Numerical results of the
		prediction intervals evaluation based on PICP, PINAW, CWC, Score,
		and AWD figures of merit}
	\label{T:pi_medidas}
	\begin{tabular}{ccccccc}
		\toprule
		Prediction Interval & PICP & PINAW & CWC ($ \kappa = 2$)&
		CWC ($\kappa = 10$)&
		Score & AWD \\
		\midrule
		BJ  &  $0.8000$ &  $0.2122 $ &  $1.4336$ &  $2.9304$ & $0.0848$ & $0.0110$  \\
		Qbeta & $1.0000$ & $0.3495 $ & $0.3495$ & $0.3495$ & $0.0699$ & $0.0000$ \\
		BPE & $0.7000$ & $0.1466 $ & $1.6384$ & $7.5357$ & $0.1498$ & $0.0631$  \\
		BCa &  $1.0000$ & $0.8994 $ & $0.8994$ & $0.8994$ & $0.1799$ & $0.0000$\\
		Block &  $0.4000$ & $0.7104 $ & $3.4287$ & $149.1236$ & $0.5611$ & $0.0916$\\
		Residual & $0.9000$ & $0.6566$ & $0.6566$ & $0.6566$ & $0.1504$ & $0.0004$ \\
		\bottomrule
	\end{tabular}
\end{table}

We also performed a sensitivity analysis based on the BCa prediction
interval considering
the methodology
proposed in~\cite{wang2018} and~\cite{espinheira2014}.
For such,
we constructed BCa prediction intervals
considering four different
number of bootstrap replications
aiming at capturing
the
effectiveness and robustness
of the prediction interval.
The sensitivity analysis
was evaluated in terms of
PICP, PINAW, CWC, Score, and AWD
figures of merit;
the results
are presented in
Table~\ref{T:pi_sens}.
We note that
the coverage rate and average length remained constant when
we increased
the number of
bootstrap
replications,
i.e.,
PICP, PINAW, CWC, Score, and AWD
show similar values
regardless
of the considered number of iterations.

\begin{table}
	\centering
	\tablesize
	\caption{
		Numerical results of the BCa prediction interval sensitivity analysis}
	\label{T:pi_sens}
	\begin{tabular}{ccccccc}
		\toprule
		Number of bootstrap iterations   & PICP & PINAW & CWC ($\kappa = 2$)
		&
		Score  & AWD\\
		\midrule
		$500$ & $1.0000$ & $0.8706$ & $0.8706$ & $0.1741$ &  $0.0000$\\
		$1000$ & $1.0000$ & $0.8994 $ & $0.8994$ & $0.1799$ & $0.0000$\\
		$2000$ &  $1.0000$ & $0.8962$ & $0.8962$ & $0.1792$ & $0.0000$\\
		$5000$ & $1.0000$ & $0.8896$ & $0.8896$ & $0.1779$ & $0.0000$\\
		\bottomrule
	\end{tabular}
\end{table}

\begin{figure}
	\centering
	\includegraphics[width=0.6\textwidth]{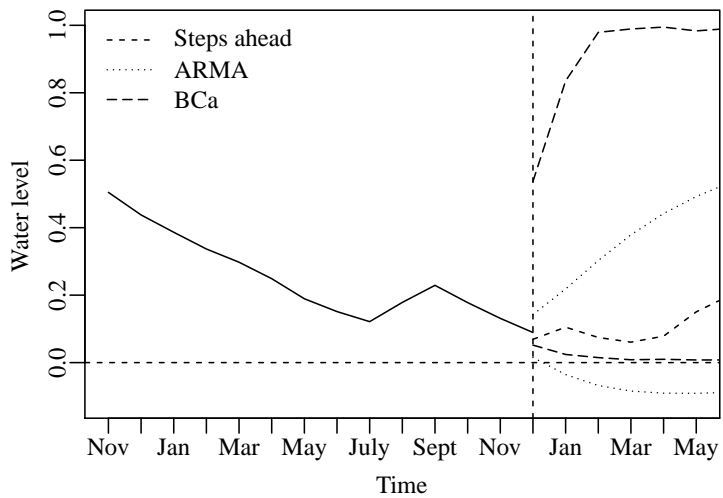}
	\caption{Prediction limits for the water levels of the Cantareira System.
	}\label{f:prevc}
\end{figure}

Finally,
we compared
BCa prediction interval
to the
prediction interval furnished
by the traditional ARMA model~\citep{Box2008}.
Figure~\ref{f:prevc} presents the last
12~observations
with the last ten original
data
values and
prediction interval
BCa
for $\alpha=0.10$.
For all observations,
the BCa prediction interval
presents
prediction limits
within the support of the
data,
while the
lower limits of the
ARMA model
prediction interval
show
values
smaller than zero.
Thus,
the results from the ARMA model
lack physical meaning
because
the
data
of interest is defined over
the
interval $(0,1)$.
This application emphasizes
the importance of
a
judicious model selection
for the construction
of reliable prediction intervals
for beta-distributed time series data.
In the ARMA model case,
the erroneous assumption of normality
led to results without clear meaning,
since they
were
outside the interval $(0,1)$.
The proposed BCa prediction interval
in $\beta$ARMA model
was identified
as the most suitable interval for this type of
data.

\section{Conclusions}
\label{s:con}

Generally,
the
ARMA models
are used for modeling and forecasting variables over time.
ARMA models
may
not be suitable
when
the variable of interest
does not satisfy normality,
as exemplified
by
variables that take
values in
the continuous interval~$(0,1)$, such as rates and proportions.
Under such conditions,
the \barma~model,
which assumes the beta distribution
to the variable of interest,
becomes a more appropriate tool.

The present work proposed
five methods for deriving
prediction intervals
under the \barma~model.
Two of the introduced methods
do not require
bootstrapping
and
are based
on the
predictions intervals for
ARMA model
and beta
quantiles distributions.
The remaining
three proposed methods
resort to bootstrapping
and stem from the
BPE, BCa, and
percentile intervals.

The prediction
intervals with bootstrapping
presented
better coverage
rate
than the
non-bootstrapping
intervals.
The BCa prediction interval
exhibited constant
values in all
scenarios considered,
with lower average length and coverage rates
close to
nominal values.
Thus, the BCa prediction interval is more reliable,
regardless of the discussed scenarios.

The proposed intervals methods
were applied to
measured
data
from
reservoir
water level
in the Metropolitan Area of S\~ao Paulo.
BCa prediction interval was
considered and compared to
the traditional ARMA models.
The limits of
the
ARMA model
prediction interval
presented
negative values
in
clear
conflict with the support of the
data
of interest.
Such mismatch
highlights the necessity of choosing an appropriate
prediction interval.
Moreover,
the
proposed
BCa prediction interval showed
values
within
the correct interval $(0,1)$.
As a conclusion,
we recommend the
use of the BCa prediction interval for constructing of
accurate prediction intervals
for
data
restricted to the interval $(0,1)$.
In future studies, we aim at addressing prediction intervals
for the \barma~model in the presence
of long dependence~\citep{pumi2019}
and seasonality~\citep{bayer2018beta},
as well as,
for competitive models that assume other distributions
for the doubly limited response variable,
such as Kumaraswamy~\citep{bayer2017kumaraswamy}
and beta binomial~\citep{palm2021}.

\section*{Acknowledgements}
	We gratefully
	acknowledge partial financial support from
	Funda\c{c}\~ao de Amparo \`a Pesquisa do Estado do Rio Grande do Sul (FAPERGS),
	Conselho Nacional de Desenvolvimento Cient\'ifico and Tecnol\'ogico (CNPq),
	and Coordena\c{c}\~ao de Aperfei\c{c}oamento de Pessoal de N\'ivel Superior (CAPES), Brazil.

\appendix

\section*{Appendix}

\label{S:apA}

In this appendix,
the numerical results for
Scenarios~\ref{i:m1},~\ref{i:m2},~\ref{i:m5},
and~\ref{i:m6} are presented
with coverage level equal to $0.90$.
Tables~\ref{t:bar04} and~\ref{t:bma04111}
show
the numerical results of simulations for
Scenarios~\ref{i:m1} and~\ref{i:m5}, respectively,
with $\mu\approx 0.4$ and
$n =100$.
Scenarios~\ref{i:m2} and~\ref{i:m6}
are found in
Tables~\ref{t:bar09} and~\ref{t:bma09},
respectively,
for
$n =100$
and $\mu\approx 0.9$.

\begin{table}
	\caption{Estimated coverage rates, average length, and balanced of the
		prediction intervals for the Scenario~\ref{i:m1} }
	\tablesize
	\centering
	\label{t:bar04}
	\begin{tabular}{lcccccccccccc}
		\toprule
		$h$ & $1$ & $2$ & $3$ & $4$ & $5$ & $6$ & $7$ & $8$ & $9$ & $10$ \\
		\midrule
		\multicolumn{11}{c}{BJ Prediction Interval}\\
		\midrule
		$\text{CR}_h$	& $0.900$ & $0.878$ & $0.883$ & $0.877$ & $0.887$ & $0.891$ & $0.885$ & $0.900$ & $0.890$ & $0.888$		\\
		$\text{A}_h$	& $0.146$ & $0.187$ & $0.188$ & $0.218$ & $0.230$ & $0.233$ & $0.246$ & $0.250$ & $0.252$ & $0.260$		\\
		$\text{CR}_h^L$	& $0.054$ & $0.062$ & $0.054$ & $0.068$ & $0.053$ & $0.054$ & $0.056$ & $0.058$ & $0.048$ & $0.057$ \\
		$\text{CR}_h^U$	& $0.046$ & $0.060$ & $0.063$ & $0.055$ & $0.060$ & $0.055$ & $0.059$ & $0.042$ & $0.062$ & $0.055$\\
		\midrule
		\multicolumn{11}{c}{Qbeta Prediction Interval}\\
		\midrule
		$\text{CR}_h$	&  $0.895$ & $0.882$ & $0.883$ & $0.884$ & $0.891$ & $0.893$ & $0.886$ & $0.899$ & $0.891$ & $0.891$	\\
		$\text{A}_h$	&	$0.146$ & $0.187$ & $0.189$ & $0.219$ & $0.231$ & $0.234$ & $0.248$ & $0.251$ & $0.254$ & $0.261$	\\
		$\text{CR}_h^L$	&	$0.054$ & $0.059$ & $0.053$ & $0.062$ & $0.047$ & $0.048$ & $0.055$ & $0.058$ & $0.043$ & $0.054$	\\
		$\text{CR}_h^U$	& $0.051$ &  $0.059$ & $0.064$ & $0.054$ & $0.062$ & $0.059$ & $0.059$ & $0.043$ & $0.066$ & $0.055$	\\
		\midrule
		\multicolumn{11}{c}{BPE Prediction Interval}\\
		\midrule
		$\text{CR}_h$	&  $0.685$ & $0.646$ & $0.687$ & $0.657$ & $0.673$ & $0.646$ & $0.672$ & $0.658$ & $0.646$ & $0.653$	\\
		$\text{A}_h$	&	$0.107$ & $0.134$ & $0.140$ & $0.163$ & $0.171$ & $0.175$ & $0.185$ & $0.187$ & $0.189$ & $0.195$	\\
		$\text{CR}_h^L$	&  $0.150$ & $0.177$ & $0.166$ & $0.160$ & $0.155$ & $0.186$ & $0.158$ & $0.168$ & $0.173$ & $0.177$		\\
		$\text{CR}_h^U$	& $0.165$ & $0.177$ & $0.147$ & $0.183$ & $0.172$ & $0.168$ & $0.170$ & $0.174$ & $0.181$ & $0.170$	\\
		\midrule
		\multicolumn{11}{c}{BCa Prediction Interval}\\
		\midrule
		$\text{CR}_h$	& $0.814$ & $0.855$ & $0.842$ & $0.862$ & $0.895$ & $0.883$ & $0.882$ & $0.903$ & $0.899$ & $0.904$	\\
		$\text{A}_h$	&	$0.142$ & $0.193$ & $0.198$ & $0.236$ & $0.251$ & $0.252$ & $0.269$ & $0.274$ & $0.277$ & $0.287$	\\
		$\text{CR}_h^L$	& $0.088$ & $0.065$ & $0.078$ & $0.070$ & $0.048$ & $0.057$ & $0.051$ & $0.053$ & $0.048$ & $0.049$		\\
		$\text{CR}_h^U$	& $0.098$ & $0.080$ & $0.080$ & $0.068$ & $0.057$ & $0.060$ & $0.067$ & $0.044$ & $0.053$ & $0.047$	\\
		\midrule
		\multicolumn{11}{c}{Block Percentile Prediction Interval}\\
		\midrule
		$\text{CR}_h$	& $0.891$ & $0.892$ & $0.873$ & $0.886$ & $0.904$ & $0.875$ & $0.879$ & $0.894$ & $0.886$ & $0.886$	\\
		$\text{A}_h$	&	$0.261$ & $0.273$ & $0.270$ & $0.271$ & $0.272$ & $0.271$ & $0.271$ & $0.271$ & $0.271$ & $0.271$	\\
		$\text{CR}_h^L$	&	$0.050$ & $0.055$ & $0.069$ & $0.059$ & $0.049$ & $0.064$ & $0.055$ & $0.058$ & $0.054$ & $0.061$	\\
		$\text{CR}_h^U$	& $0.059$ & $0.053$ & $0.058$ & $0.055$ & $0.047$ & $0.061$ & $0.066$ & $0.048$ & $0.060$ & $0.053$	\\
		\midrule
		\multicolumn{11}{c}{Residual Percentile Prediction Interval}\\
		\midrule
		$\text{CR}_h$	& $0.929$ & $0.917$ & $0.896$ & $0.882$ & $0.910$ & $0.884$ & $0.877$ & $0.898$ & $0.888$ & $0.880$ 	\\
		$\text{A}_h$	&	$0.247$ & $0.251$ & $0.263$ & $0.269$ & $0.269$ & $0.272$ & $0.272$ & $0.273$ & $0.273$ & $0.273$	\\
		$\text{CR}_h^L$	&	 $0.032$ & $0.037$ & $0.057$ & $0.062$ & $0.043$ & $0.059$ & $0.055$ & $0.056$ & $0.053$ & $0.062$	\\
		$\text{CR}_h^U$	& $0.039$ & $0.046$ & $0.047$ & $0.056$ & $0.047$ & $0.057$ & $0.068$ & $0.046$ & $0.059$ & $0.058$	\\
		\bottomrule
	\end{tabular}
\end{table}

\begin{table}
	\caption{Estimated coverage rates, average length, and balanced of the
		prediction intervals for the Scenario~\ref{i:m5} }
	\tablesize
	\centering
	\label{t:bma04111}
	\begin{tabular}{lcccccccccccc}
		\toprule
		$h$ & $1$ & $2$ & $3$ & $4$ & $5$ & $6$ & $7$ & $8$ & $9$ & $10$ \\
		\midrule
		\multicolumn{11}{c}{BJ Prediction Interval}\\
		\midrule
		$\text{CR}_h$	&	$0.854$ & $0.880$ & $0.846$ & $0.845$ & $0.847$ & $0.841$ & $0.820$ & $0.875$ & $0.847$ & $0.832$	\\
		$\text{A}_h$	&	 $0.182$ & $0.183$ & $0.183$ & $0.183$ & $0.183$ & $0.183$ & $0.183$ & $0.183$ & $0.183$ & $0.183$	\\
		$\text{CR}_h^L$	&	$0.098$ & $0.067$ & $0.083$ & $0.097$ & $0.081$ & $0.085$ & $0.104$ & $0.064$ & $0.087$ & $0.095$	\\
		$\text{CR}_h^U$	&	$0.048$ & $0.053$ & $0.071$ & $0.058$ & $0.072$ & $0.074$ & $0.076$ & $0.061$ & $0.066$ & $0.073$	\\
		\midrule
		\multicolumn{11}{c}{Qbeta Prediction Interval}\\
		\midrule
		$\text{CR}_h$	&	$0.918$ & $0.919$ & $0.902$ & $0.907$ & $0.902$ & $0.898$ & $0.888$ & $0.918$ & $0.905$ & $0.887$	\\
		$\text{A}_h$	&	$0.208$ & $0.211$ & $0.211$ & $0.211$ & $0.211$ & $0.211$ & $0.211$ & $0.211$ & $0.211$ & $0.211$	\\
		$\text{CR}_h^L$	&	 $0.052$ & $0.040$ & $0.050$ & $0.050$ & $0.040$ & $0.049$ & $0.053$ & $0.042$ & $0.045$ & $0.055$	\\
		$\text{CR}_h^U$	& $0.030$ & $0.041$ & $0.048$ & $0.043$ & $0.058$ & $0.053$ & $0.059$ & $0.040$ & $0.050$ & $0.058$		\\
		\midrule
		\multicolumn{11}{c}{BPE Prediction Interval}\\
		\midrule
		$\text{CR}_h$	&	$0.828$ & $0.835$ & $0.840$ & $0.837$ & $0.836$ & $0.840$ & $0.820$ & $0.855$ & $0.851$ & $0.835$	\\
		$\text{A}_h$	&	$0.191$ & $0.193$ & $0.195$ & $0.194$ & $0.194$ & $0.194$ & $0.194$ & $0.194$ & $0.194$ & $0.194$	\\
		$\text{CR}_h^L$	&	$0.100$ & $0.076$ & $0.081$ & $0.083$ & $0.077$ & $0.074$ & $0.093$ & $0.078$ & $0.075$ & $0.081$	\\
		$\text{CR}_h^U$	&	$0.072$ & $0.089$ & $0.079$ & $0.080$ & $0.087$ & $0.086$ & $0.087$ & $0.067$ & $0.074$ & $0.084$	\\
	    \midrule
		\multicolumn{11}{c}{BCa Prediction Interval}\\
		\midrule
		$\text{CR}_h$	&	$0.854$ & $0.884$ & $0.899$ & $0.910$ & $0.896$ & $0.900$ & $0.890$ & $0.909$ & $0.906$ & $0.888$	\\
		$\text{A}_h$	&	$0.199$ & $0.203$ & $0.212$ & $0.212$ & $0.212$ & $0.212$ & $0.212$ & $0.212$ & $0.212$ & $0.212$	\\
		$\text{CR}_h^L$	&	 $0.079$ & $0.051$ & $0.044$ & $0.045$ & $0.041$ & $0.046$ & $0.051$ & $0.042$ & $0.039$ & $0.052$	\\
		$\text{CR}_h^U$	&	$0.067$ & $0.065$ & $0.057$ & $0.045$ & $0.063$ & $0.054$ & $0.059$ & $0.049$ & $0.055$ & $0.060$	\\
		\midrule
		\multicolumn{11}{c}{Block Percentile Prediction Interval}\\
		\midrule
		$\text{CR}_h$	&	 $0.860$ & $0.875$ & $0.896$ & $0.907$ & $0.896$ & $0.898$ & $0.885$ & $0.916$ & $0.906$ & $0.880$	\\
		$\text{A}_h$	&	$0.198$ & $0.200$ & $0.209$ & $0.208$ & $0.208$ & $0.208$ & $0.209$ & $0.209$ & $0.208$ & $0.209$	\\
		$\text{CR}_h^L$	&	$0.078$ & $0.059$ & $0.053$ & $0.053$ & $0.042$ & $0.050$ & $0.060$ & $0.045$ & $0.045$ & $0.065$    \\
		$\text{CR}_h^U$	&	$0.062$ & $0.066$ & $0.051$ & $0.040$ & $0.062$ & $0.052$ & $0.055$ & $0.039$ & $0.049$ & $0.055$	\\
		\midrule
		\multicolumn{11}{c}{Residual Percentile Prediction Interval}\\
		\midrule
		$\text{CR}_h$	&	 $0.881$ & $0.882$ & $0.898$ & $0.900$ & $0.895$ & $0.901$ & $0.884$ & $0.920$ & $0.901$ & $0.883$	\\
		$\text{A}_h$	&	$0.206$ & $0.2084$ & $0.209$ & $0.209$ & $0.209$ & $0.210$ & $0.209$ & $0.209$ & $0.210$ & $0.210$	\\
		$\text{CR}_h^L$	&	$0.069$ & $0.060$ & $0.049$ & $0.057$ & $0.043$ & $0.050$ & $0.061$ & $0.044$ & $0.048$ & $0.061$	\\
		$\text{CR}_h^U$	&	  $0.050$ & $0.058$ & $0.053$ & $0.043$ & $0.062$ & $0.049$ & $0.055$ & $0.036$ & $0.051$ & $0.056$	\\
		\bottomrule
	\end{tabular}
\end{table}

\begin{table}
	\caption{Estimated coverage rates, average length, and balanced of the
		prediction intervals for the Scenario~\ref{i:m2} }
	\tablesize
	\centering
	\label{t:bar09}
	\begin{tabular}{lcccccccccccc}
		\toprule
		$h$ & $1$ & $2$ & $3$ & $4$ & $5$ & $6$ & $7$ & $8$ & $9$ & $10$ \\
		\midrule
		\multicolumn{11}{c}{BJ Prediction Interval}\\
		\midrule
		$\text{CR}_h$	&	$0.459$ & $0.468$ & $0.487$ & $0.493$ & $0.470$ & $0.458$ & $0.490$ & $0.475$ & $0.481$ & $0.459$	\\
		$\text{A}_h$	&	$0.035$ & $0.037$ & $0.039$ & $0.040$ & $0.041$ & $0.041$ & $0.042$ & $0.042$ & $0.042$ & $0.042$	\\
		$\text{CR}_h^L$	&	$0.250$ & $0.231$ & $0.224$ & $0.207$ & $0.210$ & $0.220$ & $0.190$ & $0.206$ & $0.207$ & $0.213$	\\
		$\text{CR}_h^U$	&	$0.291$ & $0.301$ & $0.289$ & $0.300$ & $0.320$ & $0.322$ & $0.320$ & $0.319$ & $0.312$ & $0.328$	\\
		\midrule
		\multicolumn{11}{c}{Qbeta Prediction Interval}\\
		\midrule
		$\text{CR}_h$	& $0.886$ & $0.886$ & $0.885$ & $0.869$ & $0.894$ & $0.859$ & $0.890$ & $0.885$ & $0.867$ & $0.860$		\\
		$\text{A}_h$	&	$0.084$ & $0.088$ & $0.094$ & $0.096$ & $0.097$ & $0.098$ & $0.098$ & $0.099$ & $0.099$ & $0.099$	\\
		$\text{CR}_h^L$	&	$0.069$ & $0.068$ & $0.060$ & $0.055$ & $0.050$ & $0.060$ & $0.037$ & $0.040$ & $0.053$ & $0.057$	\\
		$\text{CR}_h^U$	&	$0.045$ & $0.046$ & $0.055$ & $0.076$ & $0.056$ & $0.081$ & $0.073$ & $0.075$ & $0.080$ & $0.083$	\\
		\midrule
		\multicolumn{11}{c}{BPE Prediction Interval}\\
		\midrule
		$\text{CR}_h$	&	$0.745$ & $0.750$ & $0.779$ & $0.775$ & $0.786$ & $0.773$ & $0.799$ & $0.814$ & $0.783$ & $0.800$	\\
		$\text{A}_h$	&	$0.076$ & $0.082$ & $0.089$ & $0.091$ & $0.092$ & $0.092$ & $0.093$ & $0.093$ & $0.093$ & $0.093$	\\
		$\text{CR}_h^L$	&	$0.068$ & $0.062$ & $0.065$ & $0.063$ & $0.068$ & $0.088$ & $0.067$ & $0.069$ & $0.074$ & $0.080$	\\
		$\text{CR}_h^U$	&	$0.187$ & $0.188$ & $0.156$ & $0.162$ & $0.146$ & $0.139$ & $0.134$ & $0.117$ & $0.143$ & $0.120$	\\
		\midrule
		\multicolumn{11}{c}{BCa Prediction Interval}\\
		\midrule
		$\text{CR}_h$	&  $0.857$ & $0.863$ & $0.876$ & $0.865$ & $0.879$ & $0.869$ & $0.882$ & $0.894$ & $0.870$ & $0.873$	\\
		$\text{A}_h$	&	$0.083$ & $0.088$ & $0.094$ & $0.096$ & $0.097$ & $0.097$ & $0.097$ & $0.098$ & $0.098$ & $0.098$	\\
		$\text{CR}_h^L$	&	$0.091$ & $0.088$ & $0.083$ & $0.076$ & $0.072$ & $0.076$ & $0.062$ & $0.061$ & $0.067$ & $0.075$	\\
		$\text{CR}_h^U$	&	$0.052$ & $0.049$ & $0.041$ & $0.059$ & $0.049$ & $0.055$ & $0.056$ & $0.045$ & $0.063$ & $0.052$	\\
		\midrule
		\multicolumn{11}{c}{Block Percentile Prediction Interval}\\
		\midrule
		$\text{CR}_h$	&	$0.906$ & $0.912$ & $0.905$ & $0.884$ & $0.909$ & $0.895$ & $0.908$ & $0.913$ & $0.894$ & $0.881$	\\
		$\text{A}_h$	&	$0.105$ & $0.106$ & $0.106$ & $0.106$ & $0.107$ & $0.106$ & $0.106$ & $0.106$ & $0.106$ & $0.106$	\\
		$\text{CR}_h^L$	&	$0.046$ & $0.044$ & $0.050$ & $0.045$ & $0.038$ & $0.044$ & $0.029$ & $0.033$ & $0.040$ & $0.055$	\\
		$\text{CR}_h^U$	&	$0.048$ & $0.044$ & $0.045$ & $0.071$ & $0.053$ & $0.061$ & $0.063$ & $0.054$ & $0.066$ & $0.064$	\\
		\midrule
		\multicolumn{11}{c}{Residual Percentile Prediction Interval}\\
		\midrule
		$\text{CR}_h$	&	$0.916$ & $0.894$ & $0.891$ & $0.873$ & $0.895$ & $0.873$ & $0.895$ & $0.901$ & $0.876$ & $0.872$	\\
		$\text{A}_h$	&	$0.098$ & $0.099$ & $0.099$ & $0.099$ & $0.099$ & $0.099$ & $0.099$ & $0.099$ & $0.099$ & $0.099$	\\
		$\text{CR}_h^L$	&	$0.050$ & $0.059$ & $0.062$ & $0.057$ & $0.052$ & $0.064$ & $0.045$ & $0.043$ & $0.056$ & $0.063$	\\
		$\text{CR}_h^U$	&	$0.034$ & $0.047$ & $0.047$ & $0.070$ & $0.053$ & $0.063$ & $0.060$ & $0.056$ & $0.068$ & $0.065$	\\
		\bottomrule
	\end{tabular}
\end{table}

\begin{table}
	\caption{Estimated coverage rates, average length, and balanced of the
		prediction intervals for the Scenario~\ref{i:m6} }
	\tablesize
	\centering
	\label{t:bma09}
	\begin{tabular}{lcccccccccccc}
		\toprule
		$h$ & $1$ & $2$ & $3$ & $4$ & $5$ & $6$ & $7$ & $8$ & $9$ & $10$ \\
		\midrule
		\multicolumn{11}{c}{BJ Prediction Interval}\\
		\midrule
		$\text{CR}_h$	&	$0.825$ & $0.820$ & $0.745$ & $0.737$ & $0.767$ & $0.761$ & $0.755$ & $0.773$ & $0.751$ & $0.755$	\\
		$\text{A}_h$	&	$0.097$ & $0.099$ & $0.099$ & $0.099$ & $0.099$ & $0.099$ & $0.099$ & $0.099$ & $0.099$ & $0.099$	\\
		$\text{CR}_h^L$	&	$0.060$ & $0.065$ & $0.105$ & $0.099$ & $0.090$ & $0.099$ & $0.091$ & $0.096$ & $0.095$ & $0.096$	\\
		$\text{CR}_h^U$	&	 $0.115$ & $0.115$ & $0.150$ & $0.164$ & $0.143$ & $0.140$ & $0.154$ & $0.131$ & $0.1544$ & $0.149$	\\
		\midrule
		\multicolumn{11}{c}{Qbeta Prediction Interval}\\
		\midrule
		$\text{CR}_h$	&	$0.955$ & $0.948$ & $0.904$ & $0.891$ & $0.907$ & $0.905$ & $0.906$ & $0.903$ & $0.905$ & $0.890$	\\
		$\text{A}_h$	&	$0.138$ & $0.140$ & $0.140$ & $0.140$ & $0.140$ & $0.140$ & $0.140$ & $0.140$ & $0.140$ & $0.140$	\\
		$\text{CR}_h^L$	&	$0.018$ & $0.025$ & $0.043$ & $0.043$ & $0.037$ & $0.044$ & $0.039$ & $0.048$ & $0.043$ & $0.047$	\\
		$\text{CR}_h^U$	&	$0.027$ & $0.027$ & $0.053$ & $0.066$ & $0.056$ & $0.051$ & $0.055$ & $0.049$ & $0.052$ & $0.063$	\\
		\midrule
		\multicolumn{11}{c}{BPE Prediction Interval}\\
		\midrule
		$\text{CR}_h$	&	$0.824$ & $0.828$ & $0.790$ & $0.784$ & $0.809$ & $0.809$ & $0.795$ & $0.801$ & $0.794$ & $0.798$	\\
		$\text{A}_h$	&	$0.119$ & $0.121$ & $0.122$ & $0.123$ & $0.122$ & $0.122$ & $0.122$ & $0.122$ & $0.122$ & $0.122$	\\
		$\text{CR}_h^L$	&	 $0.085$ & $0.089$ & $0.099$ & $0.077$ & $0.083$ & $0.088$ & $0.090$ & $0.095$ & $0.089$ & $0.095$	\\
		$\text{CR}_h^U$	&	$0.091$ & $0.083$ & $0.111$ & $0.139$ & $0.108$ & $0.103$ & $0.115$ & $0.104$ & $0.117$ & $0.107$	\\
		\midrule
		\multicolumn{11}{c}{BCa Prediction Interval}\\
		\midrule
		$\text{CR}_h$	&	$0.883$ & $0.883$ & $0.901$ & $0.882$ & $0.907$ & $0.893$ & $0.898$ & $0.888$ & $0.898$ & $0.880$	\\
		$\text{A}_h$	&	$0.124$ & $0.129$ & $0.137$ & $0.137$ & $0.137$ & $0.137$ & $0.137$ & $0.137$ & $0.137$ & $0.137$	\\
		$\text{CR}_h^L$	& $0.061$ & $0.064$ & $0.048$ & $0.058$ & $0.047$ & $0.060$ & $0.051$ & $0.061$ & $0.053$ & $0.061$		\\
		$\text{CR}_h^U$	&	$0.056$ & $0.053$ & $0.051$ & $0.060$ & $0.046$ & $0.047$ & $0.051$ & $0.051$ & $0.049$ & $0.059$	\\
		\midrule
		\multicolumn{11}{c}{Block Percentile Prediction Interval}\\
		\midrule
		$\text{CR}_h$	&	 $0.898$ & $0.890$ & $0.899$ & $0.885$ & $0.902$ & $0.902$ & $0.897$ & $0.895$ & $0.897$ & $0.889$	\\
		$\text{A}_h$	&	$0.128$ & $0.132$ & $0.137$ & $0.137$ & $0.137$ & $0.137$ & $0.137$ & $0.137$ & $0.137$ & $0.138$	\\
		$\text{CR}_h^L$	&	$0.041$ & $0.051$ & $0.046$ & $0.046$ & $0.041$ & $0.046$ & $0.045$ & $0.050$ & $0.049$ & $0.053$	\\
		$\text{CR}_h^U$	&	$0.061$ & $0.059$ & $0.055$ & $0.069$ & $0.057$ & $0.052$ & $0.058$ & $0.055$ & $0.054$ & $0.058$	\\
		\midrule
		\multicolumn{11}{c}{Residual Percentile Prediction Interval}\\
		\midrule
		$\text{CR}_h$	&	$0.914$ & $0.900$ & $0.898$ & $0.878$ & $0.903$ & $0.893$ & $0.900$ & $0.898$ & $0.897$ & $0.888$	\\
		$\text{A}_h$	&	$0.135$ & $0.137$ & $0.137$ & $0.137$ & $0.137$ & $0.137$ & $0.137$ & $0.1374$ & $0.137$ & $0.137$	\\
		$\text{CR}_h^L$	& $0.037$ & $0.049$ & $0.048$ & $0.052$ & $0.038$ & $0.049$ & $0.041$ & $0.049$ & $0.048$ & $0.049$		\\
		$\text{CR}_h^U$	&	$0.049$ & $0.051$ & $0.054$ & $0.070$ & $0.059$ & $0.058$ & $0.059$ & $0.053$ & $0.055$ & $0.063$	\\
		\bottomrule
	\end{tabular}
\end{table}

{\small
\singlespacing
\bibliographystyle{arxiv}
\bibliography{betareg}
}

\end{document}